\documentclass[useAMS,usenatbib]{mn2e} 
\usepackage[]{graphicx}
\usepackage[fleqn]{amsmath}
\usepackage{bm}
\usepackage{natbib}
\usepackage{textcomp}
\usepackage{url}
\usepackage{xcolor}
\usepackage{hyperref}
\usepackage{color}
\usepackage{soul}
\usepackage{tabularx}
\usepackage{xr}
\usepackage{breqn}

\definecolor{refcol}{rgb}{0.3,0.,0.4}
\definecolor{refcol}{rgb}{0.,0.,0.}

\hypersetup{
  colorlinks=true,
  citecolor=refcol,
  filecolor=refcol,
  linkcolor=refcol,
  urlcolor=refcol,
  pdfborder=0 0 0.2
}

\title[GeMS astrometry performance]{Astrometric performance of the Gemini multi-conjugate adaptive optics system in crowded fields}

\author[B.~Neichel]
{\parbox{\textwidth}{Benoit Neichel,$^{1,2}$\thanks{E-mail: \texttt{benoit.neichel@lam.fr}}
Jessica R. Lu,$^{3}$
Fran\c{c}ois Rigaut,$^{4}$
S. Mark Ammons,$^{5}$
Eleazar R. Carrasco,$^{2}$
Emmanuel Lassalle$^{6}$}\vspace{0.4cm}\\
\parbox{\textwidth}{
$^{1}$Aix Marseille Universit\'e, CNRS, LAM (Laboratoire d'Astrophysique de Marseille) UMR 7326, 13388, Marseille, France\\
$^{2}$Gemini Observatory, c/o AURA, Casilla 603, La Serena, Chile\\
$^{3}$Institute for Astronomy, University of Hawaii, Honolulu, HI, USA\\
$^{4}$The Australian National University, RSAA, Mount Stromlo Observatory, Cotter Road, Weston ACT 2611, Australia\\
$^{5}$Lawrence Livermore National Laboratory Physics Division L-210 7000 East Ave., Livermore, CA 94550, USA\\
$^{6}$Ecole Centrale de Marseille, 38 rue Fr\'ed\'eric Joliot-Curie, 13013, Marseille, France \\
}}

\date{Released 2014 Xxxxx XX}

\pagerange{\pageref{firstpage2}--\pageref{lastpage2}} \pubyear{2014}

\def\LaTeX{L\kern-.36em\raise.3ex\hbox{a}\kern-.15em
    T\kern-.1667em\lower.7ex\hbox{E}\kern-.125emX}
 
\begin{document}

\label{firstpage2}

\maketitle

\begin{abstract}
The Gemini Multi-conjugate adaptive optics System (GeMS) is a facility instrument for the Gemini-South telescope. It delivers uniform, near-diffraction-limited image quality at near-infrared wavelengths over a 2 arcminute field of view. Together with the Gemini South Adaptive Optics Imager (GSAOI), a near-infrared wide field camera, GeMS/GSAOI's combination of high spatial resolution and a large field of view will make it a premier facility for precision astrometry. Potential astrometric science cases cover a broad range of topics including exo-planets, star formation, stellar evolution, star clusters, nearby galaxies, black holes and neutron stars, and the Galactic center. In this paper, we assess the astrometric performance and limitations of GeMS/GSAOI. In particular, we analyze deep, mono-epoch images, multi-epoch data and distortion calibration. We find that for single-epoch, un-dithered data, an astrometric error below 0.2 mas can be achieved for exposure times exceeding one minute, provided enough stars are available to remove high-order distortions. We show however that such performance is not reproducible for multi-epoch observations, and an additional systematic error of $\sim$0.4 mas is evidenced. This systematic multi-epoch error is the dominant error term in the GeMS/GSAOI astrometric error budget, and it is thought to be due to time-variable distortion induced by gravity flexure.
\end{abstract}


\section{Introduction}
\label{sec:introduction}

Adaptive Optics (AO) systems compensate in real-time for dynamic aberrations introduced by the propagation of light through a turbulent medium. For astronomical telescopes, AO overcomes the natural ``seeing'' limit imposed by the Earth’s atmosphere, which typically blurs images to a resolution of 0\farcs5 - 1\farcs0. This is the same resolution as a 10-50 cm telescope and is an order of magnitude worse than the diffraction limit of large 8-10 m class telescopes. Classical AO systems rely on a single natural guide star (NGS) or laser guide star (LGS) to sense the wave-front aberrations and a single deformable mirror to rapidly correct them and produce a diffraction limited science image. Most 8-10 m telescopes are now equipped with classical, ``single-conjugate'' adaptive optics (SCAO) systems.

At infrared (IR) wavelengths, ground-based AO systems deliver the highest spatial resolution and, as a result, AO can potentially deliver the best relative astrometric precision. Several groups have successfully used AO astrometry in a variety of science cases. For example, AO astrometry has been critical for studies of stars orbiting the supermassive black hole at the Galactic Center \citep{genzel2003, ghez2008gc, lu2009, Gillessen2009, fritz2010, yelda2014astrometry}. For this science case, the Keck Galactic Center studies have demonstrated astrometric uncertainties as small as $\sim$150 $\mu$as, over Fields Of View (FoV) of 10\arcsec to 20\arcsec, and repeatable over several years of observations. Similarly, \cite{cameron2009}, using an optimal weighting method demonstrated a repeatability of $\sim$ 100 $\mu$as over a two month baseline with the Palomar 5-meter AO system. Another example science case is the study of massive, young star clusters in the Milky Way to search for variations in the initial mass function and constrain models of star and cluster formation. In this case, precise proper motions are essential to distinguish cluster members from contaminating field stars \citep{stolte2008, clarkson2012, Rochau2010, kud2012}. Most of the current work is focused on the cores of the clusters since the cluster extent greatly exceeds the field of view for SCAO systems (10\arcsec-20\arcsec).  The study of star clusters and even the Galactic Center would then benefit tremendously from a wider-field AO system that delivers high spatial resolution and high-precision astrometry.\\

Astronomical observations with SCAO can only be obtained in the vicinity of relatively bright stars (R $\sim$ 15). This puts a severe restriction on performance, limiting the fraction of the sky accessible to only about 5 per cent. On the other hand, the corrected field is limited to a few tens of arc-seconds due to anisoplanatism. Multi-Conjuagte AO (MCAO) was first theorized and later developed in detail to overcome these limitations (e.g. \cite{beckers1988increasing, johnston1994analysis, ellerbroek1994first, miska2001tomo}). By using multiple Laser Guide Stars (LGSs - e.g. \cite{tallon1990adaptive, fried1994analysis}), MCAO systems can potentially deliver AO correction over an area ten to twenty times larger than what was possible with previous AO systems.\\

The Gemini Multi-Conjugate Adaptive Optics Systems (a.k..a GeMS) is the first LGS-MCAO systems offered to the community \citep{rigaut2014gems,neichel2014gems}. It uses five LGSs distributed on a 1 arcmin constellation to measure and compensate for atmospheric distortions and delivers a uniform, close to diffraction-limited Near-Infrared (NIR) image over an extended FoV of 2 arcminutes. The GeMS's LGSs are produced by a 50W laser split into 5 distinct 10-Watt beacons by a series of beamsplitters. The MCAO correction is performed by two Deformable Mirrors (DMs) conjugated to 0 and 9 km (hereafter DM0 and DM9 respectively) and one Tip-Tilt Mirror (TTM). After this, a first dichroic beam splitter is responsible for separating the visible from NIR light, sending the former to the WFSs, and the latter to the science output to feed the instruments. At the GeMS output, the corrected beam can be steered towards different science instruments attached to the Cassegrain focus instrument cluster. The main instrument used to date is Gemini South Adaptive Optics Imager (GSAOI; \cite{mcgregor2004gemini}), a 4k x 4k NIR imager covering 85 arcsec x 85 arcsec designed to work at the diffraction limit of the 8-m telescope.\\

In the literature, much attention has been paid to astrometry with MCAO systems (e.g. \cite{trippe2010,Meyer2011,schoeck2013tmt}). Improvement of the PSF width decreases the astrometric error due to photon noise, so MCAO should improve the overall astrometric error budget. But MCAO systems are also capable of inducing field distortions through deformable mirrors conjugated to higher altitude layers. For instance, a stuck or broken DM actuator at 9 km altitude will induce local plate scale distortions that will produce additional systematic errors. \cite{Meyer2011} performed an analysis of the astrometric performance delivered by MAD, an MCAO demonstrator developed by ESO and temporarily installed and tested at the ESO/VLT in 2007 \citep{marchetti2007mad}. They analyzed two globular cluster, and found a precision around $\sim$ 1 mas for stars corresponding to 2MASS K magnitudes between 9 and 12.  This performance was lower than expected, and the authors attributed the degradation to frame dithers that introduce additional distortions. More recently, \cite{rigaut2012gems} performed a preliminary analysis of images obtained with GeMS, and demonstrated a precision down to $\sim$0.4 mas for single epoch data. This result was latter confirmed by \cite{ammons2013astrometry} for single-epoch, sparse field observations.\\

This paper presents an evaluation of the astrometric performance delivered by GeMS. In particular, we analyze deep, single-epoch images, multi-epoch data, dithered data and distortion calibration. This paper only considers the analysis of crowded fields, with densities higher than 30 stars per arcmin$^2$ (see Sec. \ref{ssec:disto} for the definition justification). A companion paper (Ammons et al. - in preparation) will be dedicated to analysis of the sparse field case.\\


The outline of this paper is the following: in section \ref{ssec:simuls} we present a set of simulations to derive the theoretical performance that one can achieve with GeMS/GSAOI, we test different algorithms to measure the star positions, and we describe the data analysis used ; in Section \ref{sec:data}, we describe the observations used to asses the GeMS/GSAOI astrometric performance ; in Section \ref{sec:results} we present the results in terms of astrometric performance over single and multi-epochs ; and finally section \ref{sec:discussion} discusses the results.

\section{Methods and simulations}
\label{ssec:simuls}

\subsection{Star position extraction}

The foundation of any astrometry program is the stellar position measurement. In order to discriminate different potential algorithms, we have performed intensive tests and compared the performance of respectively $Sextractor$ \citep{bertin1996}, $StarFinder$ \citep{diolaiti2000} and $Yorick$ \citep{munro1995yorick}. In this work, we did not try to use DAOPHOT \citep{stetson1987}, which is a widely used photometry and astrometry package integrated within $Iraf$. This choice was motivated by the results of \cite{diolaiti2000}, who demonstrated that $StarFinder$ provides comparable results in terms of photometry and astrometry as DAOPHOT, but also because the optimization of DAOPHOT is not trivial, as shown by \cite{schodel2010}.\\

$Sextractor$ is a well known tool, widely used by the astronomical community, especially to build catalogs of large scale galaxy-survey data. $Sextractor$ builds a PSF model from the data, using the package $PSFex$. 
The generated PSF models can then be used to find and fit stars and extract their photometry and astrometry. It is important to note that $Sextractor$ has not been designed to perform accurate astrometry; however, the package is worth testing since it is fast and accurate PSF models over the field are a key ingrediant for astrometry.\\

$StarFinder$ is an IDL-based software package developed for PSF-fitting to extract astrometry and photometry in AO images of stellar fields. $StarFinder$ is currently one of the most used tools for AO-based astrometry studies. $StarFinder$ builds a PSF model directly from the data by analyzing a set of PSF stars selected by the user. 
The stellar astrometry and photometry is then extracted by correlating the PSF model with the data. \\

$Yorick$ is an interpreted-language. It does not provide a specific tool for astrometric measurement, but we used it to develop a fitting method to measure stellar positions. We fit the star intensity distribution using a Moffat profile defined by:
\begin{equation}
I=I_0*\left[ 1+(X/dx)^2+(Y/dy)^2 \right]^{-\beta} + I_{bkg}
\end{equation}
where $X=(x-x_0)\cos\theta + (y-y_0)\sin\theta$ and $Y=(y-y_0)\cos\theta - (x-x_0)\sin\theta$. The free parameters of the fit are the positions ($x_0$ and $y_0$), the intensity at the center ($I_0$), the width in both directions ($dx$ and $dy$), the position angle ($\theta$), and the beta index ($\beta$). The background ($I_{bkg}$) is fit simultaneously. \\

\begin{figure}
  \caption[] {Performance of different algorithms used to extract star positions for different flux conditions. $Sextractor$ is in red, $StarFinder$ in magenta and the $Yorick$ fitting method is in black. Flux is measured from the $Yorick$ fitting procedure. The PSF FWHM is 4 pixels.}
  \includegraphics[width = 1.0\linewidth]{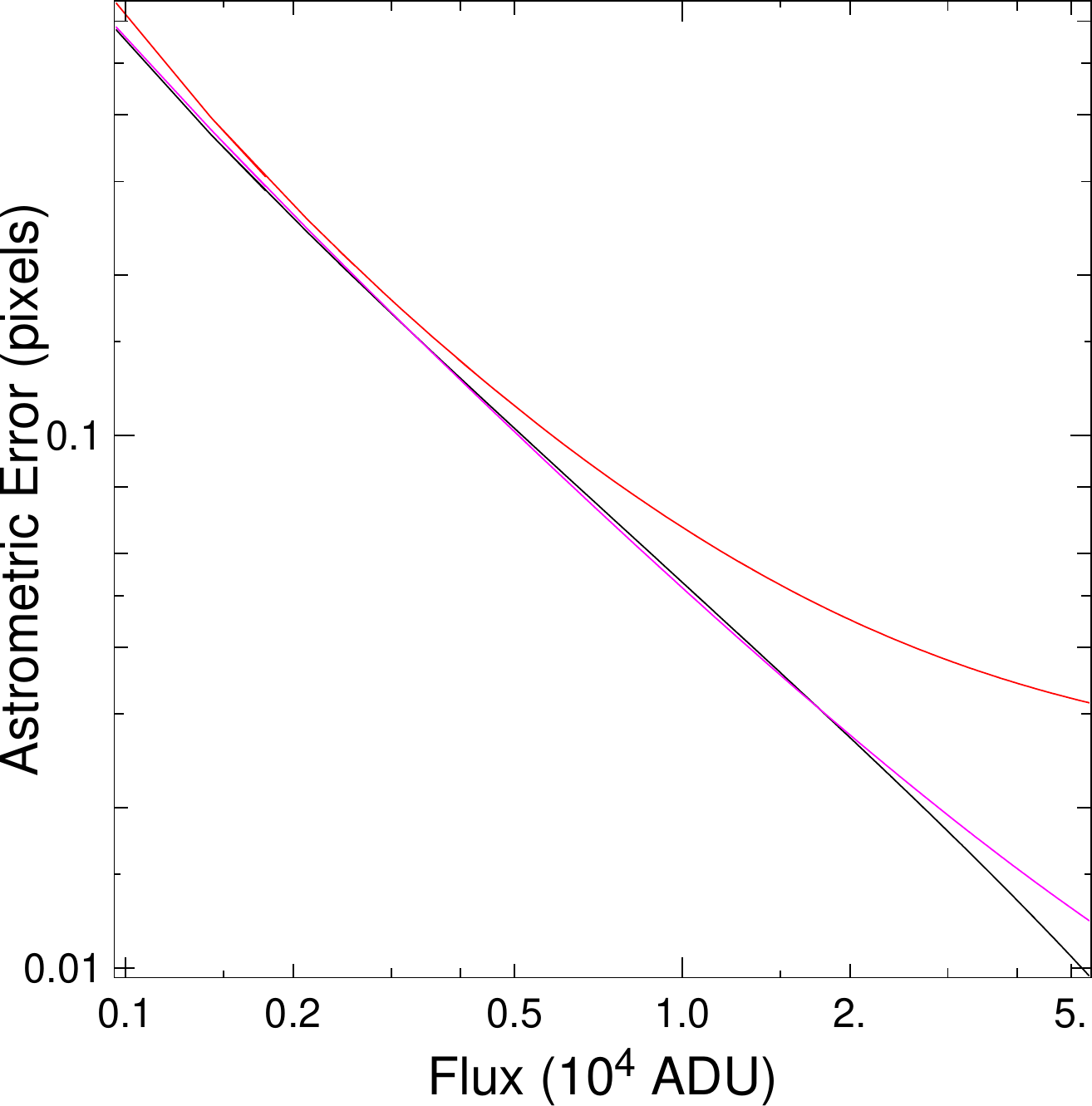}
  \label{fig:yorvsstarf}
\end{figure} 

We have compared the performance of these three algorithms both on simulated and real images. Results on simulated images are shown in Fig. \ref{fig:yorvsstarf}. Details on how the simulated images have been built can be found in Sect. \ref{sec:simmul}. The PSF FWHM is 4 pixels. As shown in Fig. \ref{fig:yorvsstarf}, $StarFinder$ and $Yorick$ perform similarly at all fluxes, but $Sextractor$ encounters a noise floor at a few hundredths of a pixel. As described above, $Sextractor$ has not been designed to derive precise astrometric measurements, and even when taking into account potential PSF variations over the field, its astrometric performance are not competitive with dedicated tools like $StarFinder$. 
For the real images, we compared the star list position measured on 3 images, by using both the $Yorick$ fitting method and $StarFinder$, this latter being run independently by J.R.L and S.M.A. Results show that, once the images are cross-registered and compensated for potential Tip-Tilt and rotation, the three methods agree on the estimation of the astrometric error within 5\%. We conclude that the choice of the star position measurement method, between $StarFinder$ and $Yorick$, does not impact the final results. In this paper, all the star positions extraction will be done with the $Yorick$ fitting method.

\subsection{Data analysis}
\label{sec:datanalysis}

Once the stellar positions have been extracted from the images, the following analysis steps are performed on single-epoch data sets. 
First, a master reference frame is built and all single-frame coordinates are later transformed to this reference frame. This reference frame is created following the method described in \cite{Meyer2011}: we use the best indiividual frame, chosen according to the highest mean Strehl Ratio (SR) in the images as the initial reference frame, and we map all the stellar positions from each individual frame onto this reference frame. Mapping individual frames to the reference frame involves adjusting translation, rotation, plate scale, and high order distortion terms to match those of the reference. Once all frames in a set have been corrected for distortions, the master-coordinate-frame is created by averaging the position of each star over all frames. 
Second, each individual frame is re-mapped to the new master coordinate system, again by compensating plate-scale and high order distortions.

After all images are aligned to a common coordinate system, we use several different methods for analyzing astrometric precision and accuracy. First, we look at the positions of an individual star and how it varies in an aligned stack of images. 
The star's position in the image stack is averaged and the root mean square error is one metric for the positional uncertainty. 
We will refer to this error as the {\em standard astrometric error} with the symbol $\sigma_{\mathrm{STD}}$.
Second, we can examine the positional difference between pairs of stars. The separation between the two stars in the image stack is averaged and the root mean square error is a second metric for positional uncertainty. This metric is useful for examining spatial dependencies in the astrometric error and we will refer to it as the {\em pairwise astrometric error} with the symbol $\sigma_{\textrm{pair}}$.

\subsection{Impact of PSFs variations}
\label{sec:simmul}

The theoretical limit of astrometric precision is defined by photon error and is given by \cite{lindegren1978} to be:
\begin{equation}
\sigma \propto \frac{FWHM}{SNR}
\end{equation}
where SNR is the signal to noise ratio determined as the ratio of flux inside a certain radius and the standard deviation of the flux inside the same area in the noise image \citep{fritz2010}. However, there are many other potential sources of errors that can affect the astrometric performance of an instrument. For instance, images taken through the earth's atmosphere suffer from positional jitter due to differential tip-tilt distortions. Time-variable distortions in the telescope or adaptive optics system can also introduce systematic astrometric errors. Finally, lack of PSF knowledge and PSF variability across the field of view can also limit the astrometric precision. For AO-assisted observations, these error terms have been described in details in e.g. \cite{fritz2010} or \cite{trippe2010}. In this section, we use simulations to evaluate the impact of PSF variations over the field. Even if MCAO provides a much more uniform correction over the field, some PSF variations remain that may affect the astrometric performance. \\

In order to reproduce realistic PSF variations over the field and with time, we use a full end-to-end Monte Carlo simulation derived from YAO\footnote{http://frigaut.github.io/yao/index.html}. This simulation tool has been designed to reproduce and analyze GeMS performance \citep{rigaut2010myst}. Therefore, it replicates all specific GeMS parameters, including the LGS and NGS constellation geometry, noise propagation statistics, etc. Based on this tool, we simulated a set of 35 PSFs at H-Band, spanning the full 85 $\times$ 85 arcsecond field. The PSFs are simulated by averaging short-exposure PSFs, computed from the residual phase maps in each PSF direction. In order isolate the impact of PSF variations, the short exposure phase screens are tip-tilt filtered, which removes the effect of differential tip-tilt jitter. This is equivalent to assume an infinitely bright constellation of NGS. Hence, the only PSF position variations are caused by PSF shape variations. These PSFs are then embedded in a simulated background image where the background flux level is derived from GSAOI on-sky H-band data. We simulate a 15s exposure, which corresponds to 300 ADU of sky-background. 
Photon noise and detector read-out noise (11.7 ADU in bright mode) are added to form the final image. Flat-fielding is assumed to be noise-free and the pixel response is assumed to be time-invariant. Different PSF flux levels are also explored by scaling the PSF before the photon-noise computation. Finally, for each PSF flux level, we simulated a set of twelve, 15 seconds exposure images, that are representative of the atmospheric time variations expected for GeMS/GSAOI observations. In parallel, we also created a set of ideal Airy PSFs that will be used to derive the fundamental astrometric performance limit. 

\begin{figure}
  \caption[] {Astrometric error vs. flux (top) and magnitude (bottom) estimated from simulations. The black dashed curves show the errors for a perfect airy function with a FWHM of 60, 80, and 100 mas respectively (from bottom to top). The blue solid curve shows the error for simulated PSFs with an average SR=23\% (FWHM = 75mas). Magenta is for SR=10\% (FWHM = 80mas). Red is for SR=6\% (FWHM = 100mas). The blue dotted line also shows the astormetric errors for the S=23\% simulation but only from PSFs within 30\arcsec $\times$ 30\arcsec.}
  \begin{tabular}{c}
  \includegraphics[width = 1.0\linewidth]{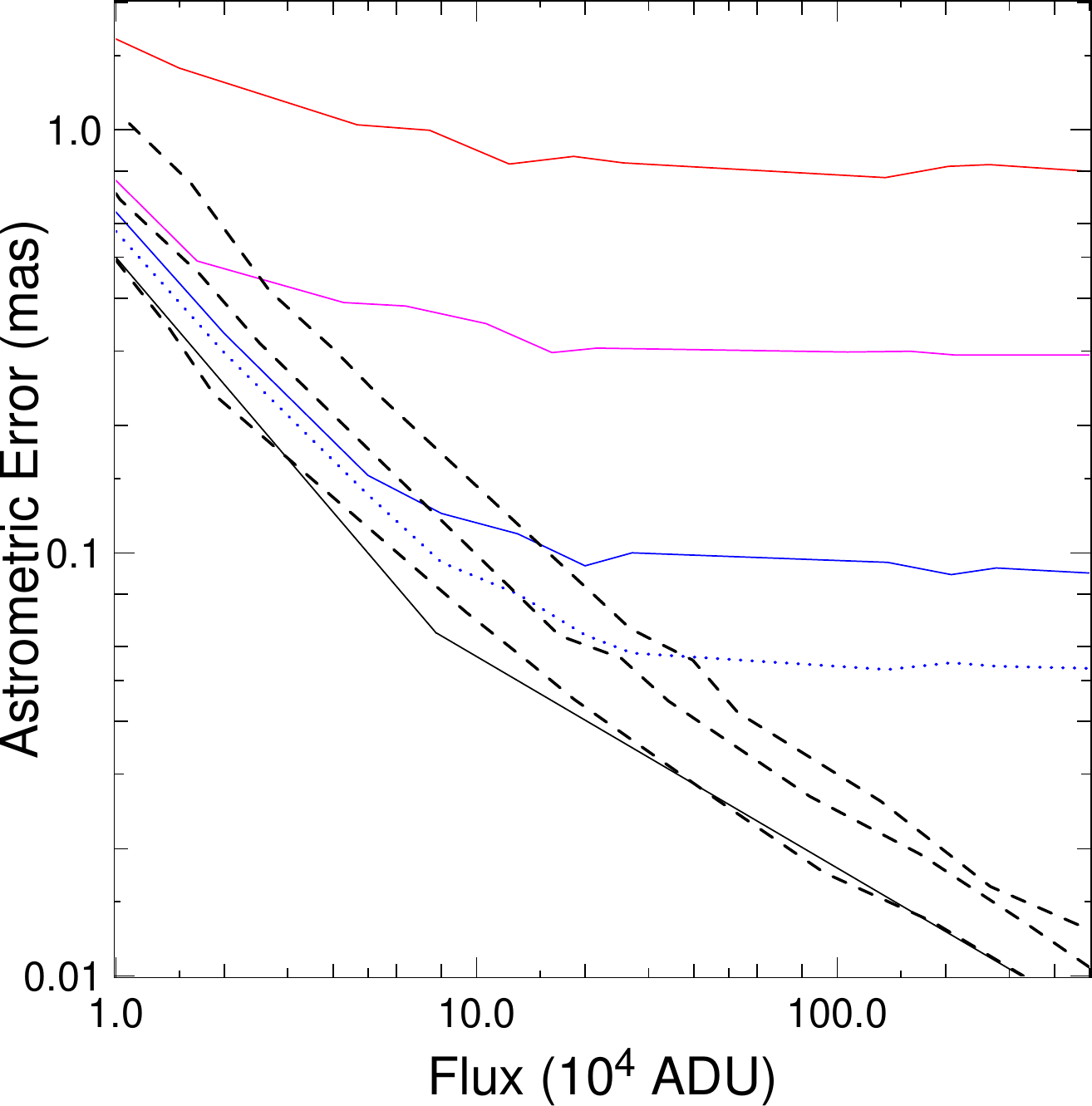} \\
  \includegraphics[width = 1.0\linewidth]{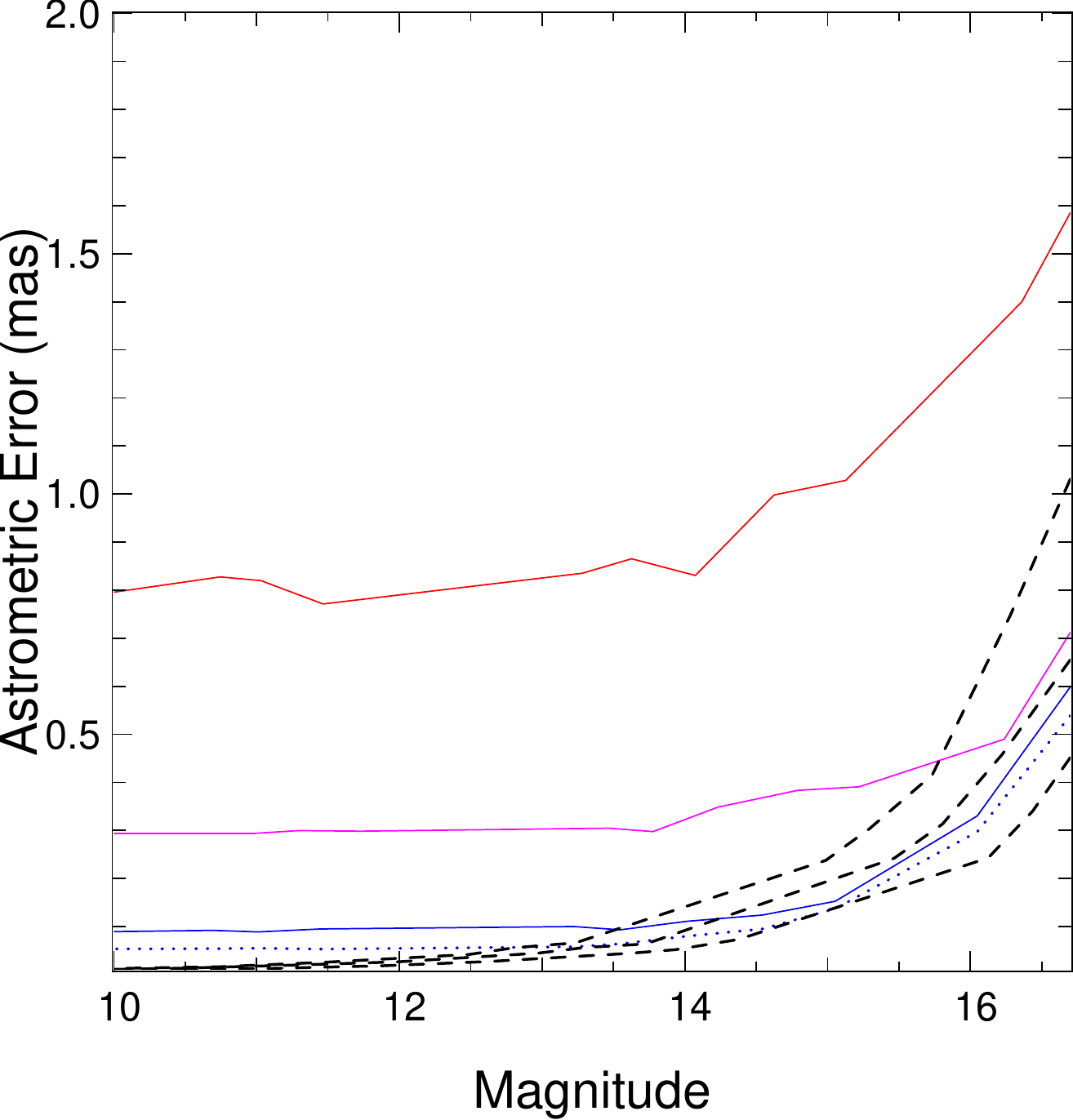}  
  \end{tabular}
  \label{fig:astro_error1}
\end{figure}

Results are presented in Fig. \ref{fig:astro_error1}. The astrometric error, computed as the RMS error of the positions across all 12 images, is given in milli-arcsec (mas), using a pixel size of 20 mas and the flux has been measured from the $Yorick$ PSF-fitting procedure. The top plot in Fig. \ref{fig:astro_error1} shows the astrometric error for different PSF fluxes, and the bottom plot uses magnitude units with a zeropoint of ZP = 16.8. This zeropoint has been calibrated against faint 2MASS stars and is accurate to $\sim$ 0.2 magnitude. The three dashed lines show the astrometric errors one can get with perfect PSFs, for 3 different FWHMs of 60, 80 and 100mas. The solid line shows a fit to the 60 mas data, that highlights two regimes: a $1/(\mbox{Flux})$ evolution for fluxes lower than 7.5 x 10$^4$ ADU (equivalent to m$_H$ $\simeq$15), and a $1/\sqrt(\mbox{Flux})$ for higher fluxes. The former regime is dominated by the detector and sky noise, the latter being dominated by the PSF photon noise. These results  are very consistent with those derived by \cite{fritz2010} (see their Fig. 2 for a detailed analysis of the different noise regimes). The red, magenta and blue  solid lines show how the astrometric error behaves when we simulated PSFs with SR=6\% (FW HM= 100 mas), 10\% (FWHM = 80 mas) and 23\% (FWHM = 75 mas) respectively. Finally, the blue dotted line shows the astrometric errors for the PSF located inside a 30\arcsec $\times$ 30\arcsec FoV.

Figure \ref{fig:astro_error1} indicates that PSF variation across the field will define the astrometric noise floor for bright stars (K $<$ 15). The PSF variability is expected to be higher for low-Strehl PSFs, so the astrometric error should be larger. Low order wavefront residuals produce asymmetric intensity patterns in the PSF, which bias the position measurement. If the low-order aberrations are not properly controlled by the MCAO system and vary across the field, or if there are quasi static aberrations (e.g. \cite{neichel2014gems}), the astrometric performance can be affected.
In that case, PSF estimation / reconstruction methods \citep{gilles2012psf,jolissaint2012psf} could potentially bring a significant gain for the astrometric performance. 

\begin{figure}
  \caption[] {Distribution of the astrometric error over the field for the SR = 23\%, FWHM = 75mas case.}
  \includegraphics[width = 1.0\linewidth]{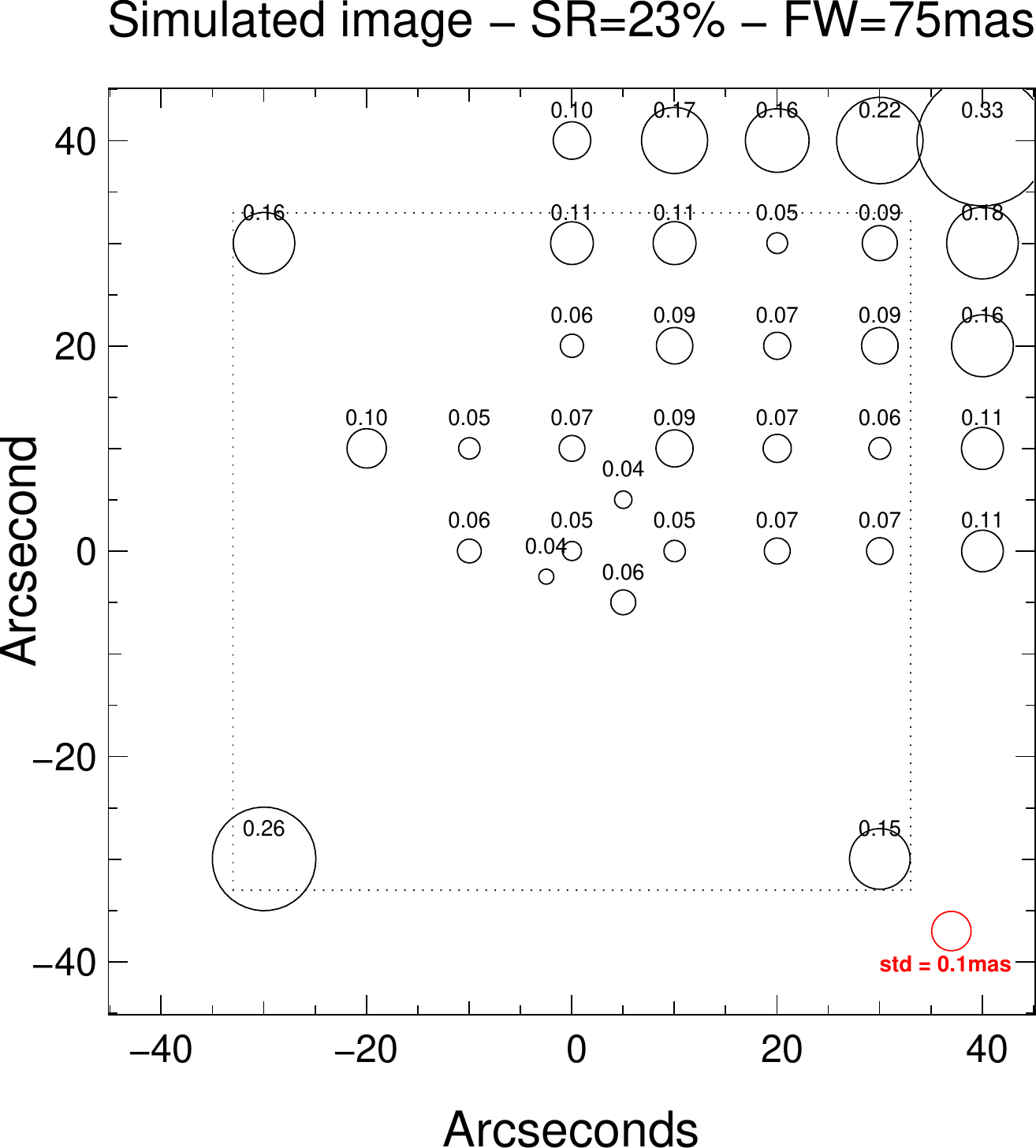}
  \label{fig:astro_error3}
\end{figure}

In Fig. \ref{fig:astro_error3}, we plot how the errors are distributed over the field, for the SR=23\% high-flux, case. 
Due to the LGS geometry, the center of the GSAOI field is better constrained than the edge of the field. If we restrict the PSFs used to estimate the astrometric error to a 30x30arcsec FoV (represented as the square dots in Fig. \ref{fig:astro_error3}), where less PSFs variations are observed, then the performance increases, as illustrated by the blue-dot line in Fig. \ref{fig:astro_error1}.

\begin{figure}
  \caption[] {Pairwise astrometric error as a function of the distance between stars for the SR = 23\%, FWHM = 75mas case (gray dots). Solid lines are the median value per 10\arcsec separation bin.}
  \includegraphics[width = 1.0\linewidth]{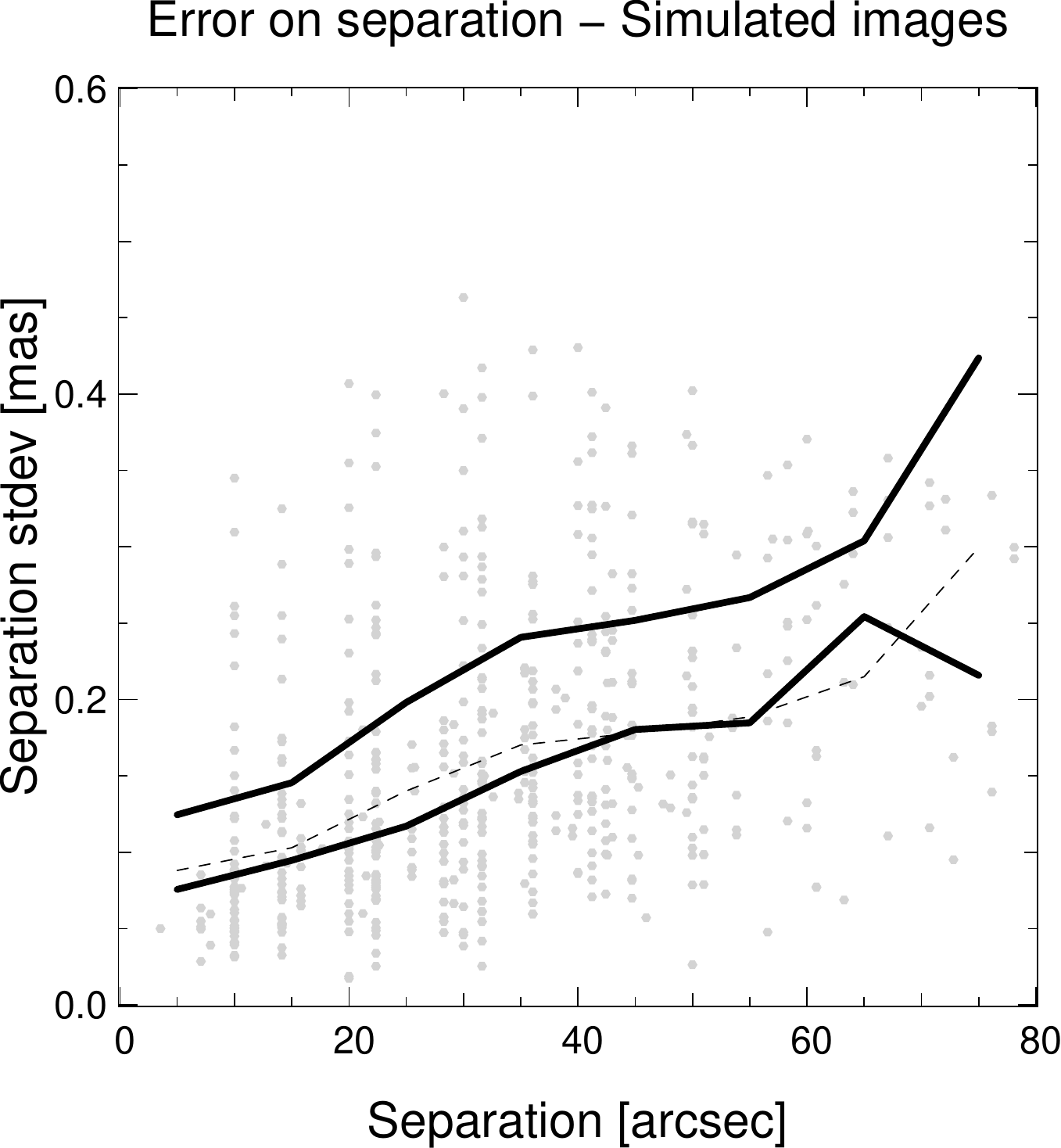}
  \label{fig:astro_error4}
\end{figure}

Figure \ref{fig:astro_error4} shows the pairwise astrometric error. The solid lines indicates the median of the data points, which are shown as gray dots, computed in bins of 10\arcsec. The upper curve is for un-binned images, in that case each of the twelve images is considered independently. For the bottom curve, images have been binned by pair, i.e. 1 with 2, 3 with 4, etc. Finally, the dashed line is a re-plot of the upper solid line, but divided by $\sqrt{2}$ as one would expect if the errors are from random sources, uncorrelated from image to image. The good agreement between the bottom solid curve, and the dashed curve indeed shows that the impact of PSF variations over the field can be treated as an additional, uncorrelated error source. The fact that the error increases with the distance can be interpreted as follows. For small separations, stars are within a given isoplanatic patch, and are all elongated on a similar manner. However, when going over large distances between the stars, the relative elongation may be different as the stars are seen through different isoplanatic patch. 

\section{Observations} 
\label{sec:data}

During the GeMS/GSAOI commissioning, several crowded stellar fields were observed in order to test the astrometric precision and accuracy of the system.
The list of targets includes the globular clusters NGC 1851 ($\alpha =$ 05 14 06.95, $\delta =$ -40 02 47.9), the open cluster NGC 2362 ($\alpha =$ 07 18 35.94, $\delta =$ -24 58 33.7), and a field in the Large Magellanic Cloud (LMC) ($\alpha$ =05 21 56.5, $\delta$ = -69 29 54.1). 
These fields have been observed either with the H-band filter ($\lambda=1.635$ \micron, $\Delta\lambda=0.290$ \micron) or with the Ks-band filter ($\lambda=2.150$ \micron, $\Delta\lambda=0.320$ \micron). 
Integration times for individual exposures varied between 5.5 s and 30 s, depending on the quality of the adaptive optics correction and other GeMS commissioning tests being conducted.

\subsection{Single epoch data}
\label{ssec:datasingle}
The first data sets considered are single epoch, un-dithered observations. In that case, multiple images are taken across the same night, with the stars always at the same pixel location. These {\em un-dithered} data sets allow us to examine the highest possible astrometric precision that could be achieved given a perfectly known distortion solution and to test the astrometric stability of the system. All the data available fulfilling these conditions are summarized in Table \ref{tab:singleepoch_data}. Table \ref{tab:singleepoch_data} gives the target name, the date, the filter used, the number of available images per data set, the integration time, and total integration time of the sample, the averaged FWHM measured over the field and along all the images, the averaged SR, the number of stars selected to compute the astrometric performance and finally a position flag. Targets with different position flags were observed at a different pixel locations. 

\begin{table*}
\caption{Single epoch - un-dithered data.}
\begin{center}
\begin{tabular}{lccccccccc}
\hline \hline
Target & Date (UT) & Filter & Texp & \# Im & Ttot & $<$FWHM$>$ & $<$SR$>$ & \# Stars & Pos \\ \hline
NGC2362 & 2011 Dec 15 & H & 15 s & 25 & 375 s & 67 mas & 15.5\% & 72 & 1\\
NGC2362 & 2011 Dec 15 & H & 15 s & 9 & 135 s & 76 mas & 14\% & 82 & 2\\
NGC2362 & 2011 Dec 18 & H & 15 s & 8 & 120 s & 63 mas & 18\% & 68 &1\\
NGC2362 & 2011 Dec 19 & H & 15 s & 27 & 405 s & 59 mas & 23\% & 73 & 3\\
NGC2362 & 2011 Dec 20 & H & 15 s & 18 & 270 s & 63 mas & 16.5\% & 73 & 3\\ 
NGC1851 & 2012 Nov 05 & H & 30 s & 39 & 1170 s & 85 mas & 7\% & 621 & 1\\ 
LMC & 2012 Dec 28 & H & 15 s & 17 & 255 s & 110 mas & 6\% & 149 & 1\\ \hline \hline
\end{tabular}
\end{center}
\label{tab:singleepoch_data}
\end{table*}

\begin{figure*}
  \caption[] {NGC2362 at H band taken with GeMS + GSAOI. The field of view is 85 \arcsec x 85 \arcsec -  the white cross is the gaps between the HAWAII 2RG arrays of GSAOI. This image is a combination of thirteen 15 seconds exposures, acquired over the course of 2 hours, during technical tests on December 19, 2011. The averaged FWHM is 60 mas and averaged SR is 23\%.}
  \includegraphics[width = 0.85\linewidth]{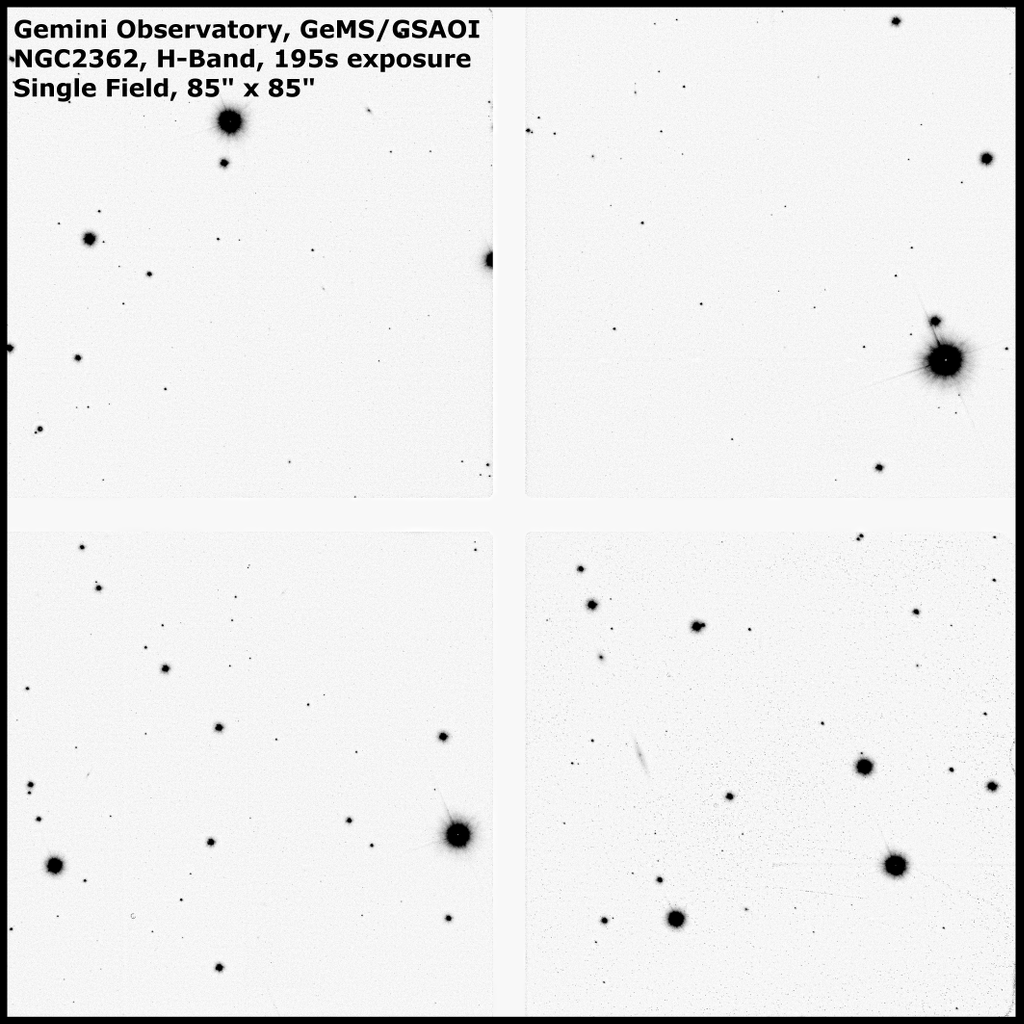}
  \label{fig:ngc2362}
\end{figure*}

\begin{figure}
  \caption[] {SR and FWHM map for one random frame of NGC2362, observed on the night of December 19th, 2011.}
  \begin{tabular}{c}
  \includegraphics[width = 1.0\linewidth]{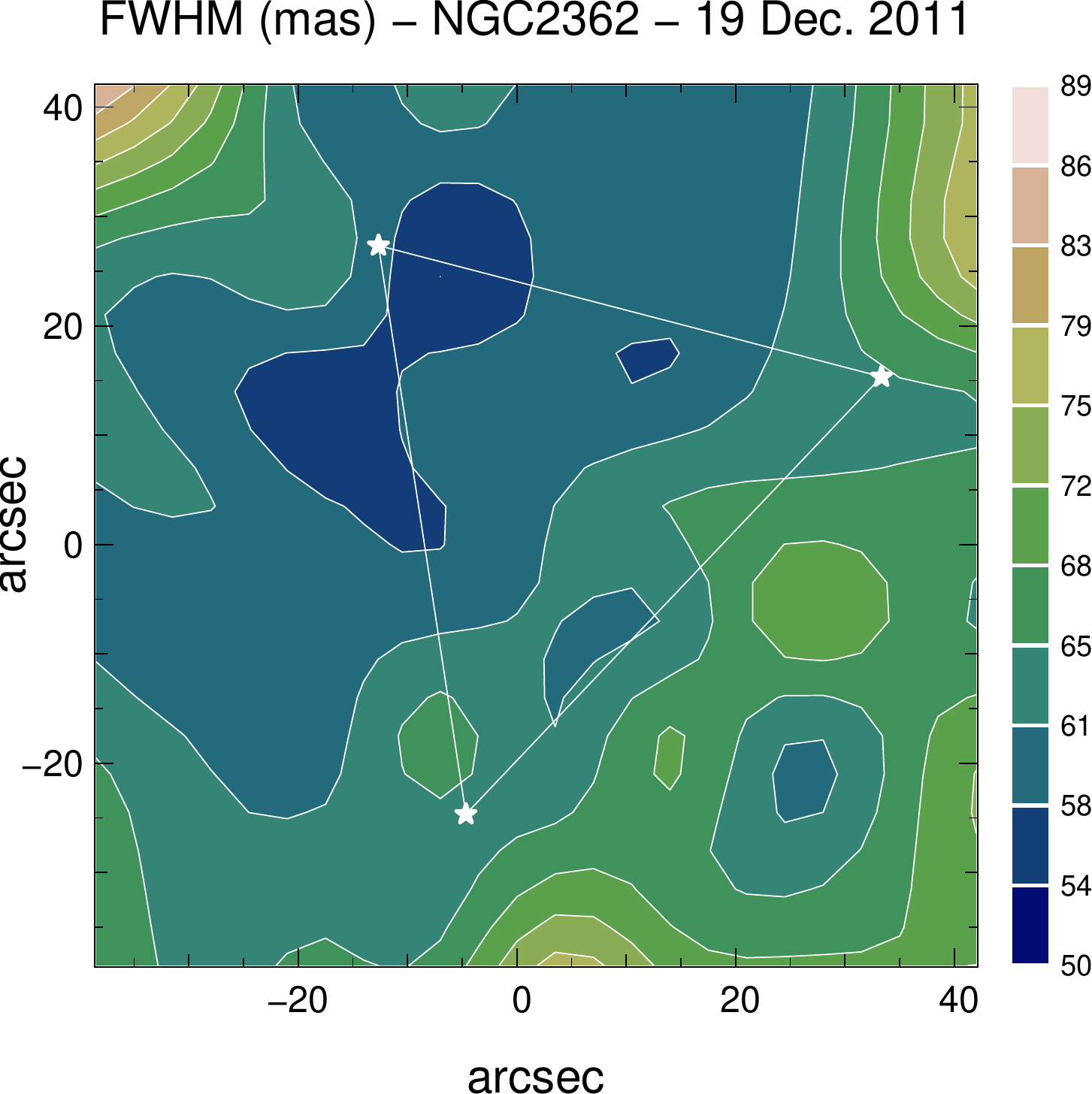} \\
  \includegraphics[width = 1.0\linewidth]{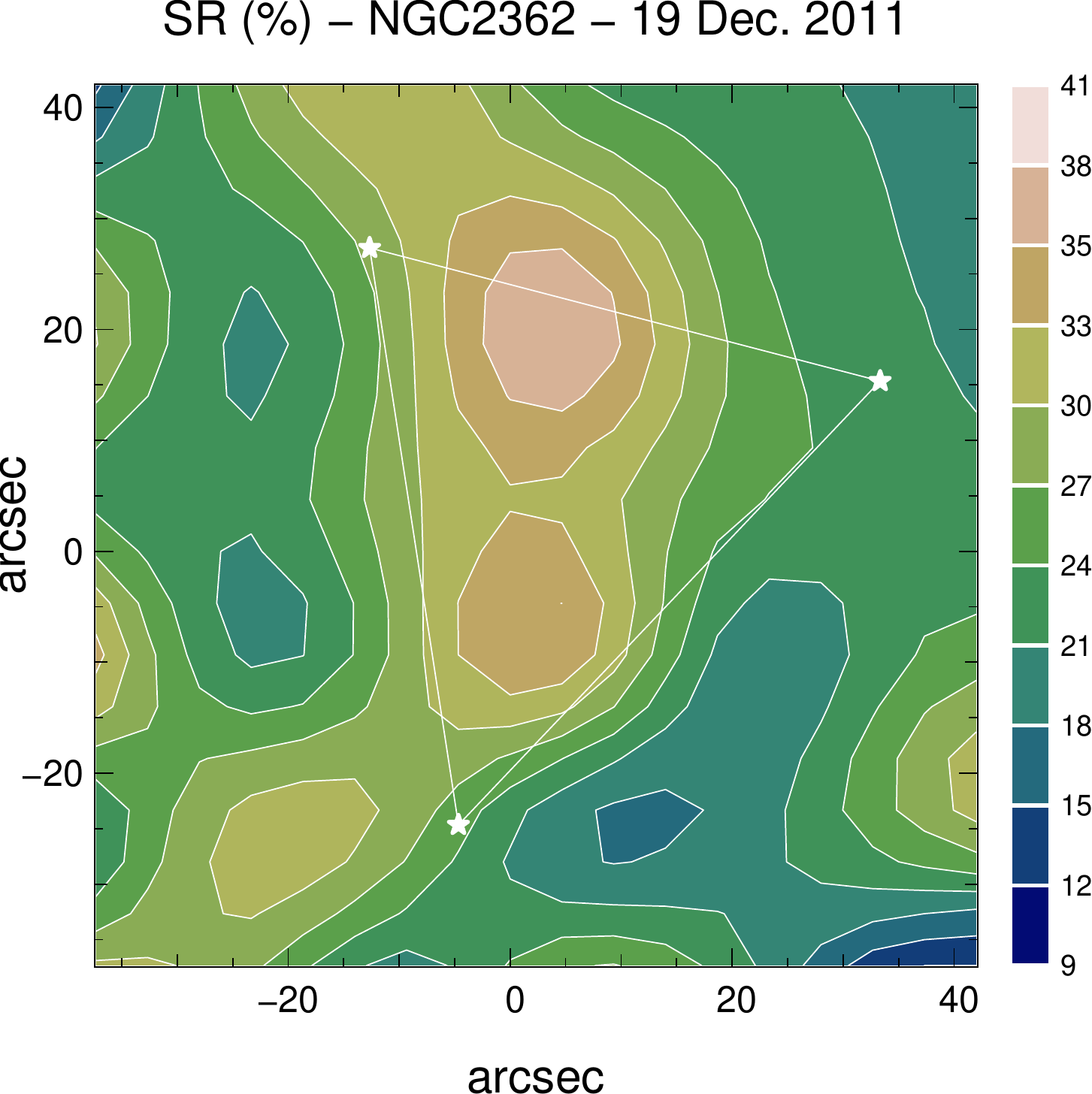}  
  \end{tabular}
  \label{fig:fwhm_map1}
\end{figure}

Fig. \ref{fig:ngc2362} shows an example of the NGC2362 field, taken on the night of December 19th, 2011. Fig. \ref{fig:fwhm_map1} shows a typical SR and FWHM map, extracted randomly from one of the images observed 2011 Dec. 19th. This illustrates the uniformity of the correction.

\subsection{Multi epoch data}
While positions can be measured very precisely in a single or short series of exposures, ultimately, it is the degree of repeatability of the measurement over nights, months, and years that is important. Ideally, the positional difference for a star observed on adjacent nights (assuming zero proper motion) would be consistent with the astrometric error in a single night.
The multi-epoch data sets include individual targets that been observed on different dates with (1) the same natural guide stars, (2) stars located on the same pixel position of the detector, and (3) no dithers. The last two points are necessary to eliminate the effects of static distortions and uncover other systematics effects introduced over different epochs. Data fulfilling these conditions are summarized in Table \ref{tab:multiepoch_data}. Note that some of the NGC2362 data presented in Table \ref{tab:singleepoch_data} can also be used as multi-epoch data. The LMC data covers the longest period of time. Indeed, this field has been used observed periodically to calibrate the World Coordinate System (WCS) solution of the GSAOI camera \citep{carrasco2012results}. 

\begin{table*}
\caption{Multi epoch - un-dithered data.}
\begin{center}
\begin{tabular}{lccccccccc}
\hline \hline
Target & Date (UT) & Filter & Texp & \# Im & Ttot & $<$FWHM$>$ & $<$SR$>$ & \# Stars & Pos \\ \hline
LMC & 2012 Dec 28 & Ks & 15 s & 4 & 60 s & 80 mas & 22.5\% & 149 & 1\\
LMC & 2012 Dec 29 & Ks & 5.5 s & 8 & 44 s & 81 mas & 22\% & 149 & 1\\
LMC & 2013 Jan 30 & Ks & 5.5 s & 8 & 44 s & 101 mas & 12.5\% & 149 & 1\\
LMC & 2013 Oct 17 & Ks & 10 s & 35 & 350 s & 91 mas & 17\% & 149 & 1\\
LMC & 2014 Feb 10 & Ks & 15 s & 11 & 165 s & 95 mas & 13\% & 149 & 1\\
NGC1851 & 2012 Dec 30 & Ks & 5.5 s & 29 & 160 s & 81 mas & 24\% & 508 & 2 \\
NGC1851 & 2012 Dec 31 & Ks & 5.5 s & 32 & 172 s & 87 mas & 18\% & 508 & 2 \\
NGC1851 & 2013 Jan 28 & Ks & 10 s & 14 & 140 s & 84 mas & 21\% & 508 & 2 \\ \hline \hline
\end{tabular}
\end{center}
\label{tab:multiepoch_data}
\end{table*}

\subsection{Dithered data}
\label{ssec:dithered data}

The third data sets considered in this study are single-epoch, but dithered data. In that case, the stars are dithered over the pixels of the detector. This is a classical way to mitigate hot and dead pixels present in NIR arrays, as well as to fill the gap lying in between the GSAOI detectors. However, if static optical distortions are present in the camera optics, dithering may impact the astrometric performance. To evaluate the impact of dithering on the astrometric performance, we will use a data set acquired on NGC2362 target, as described in Table \ref{tab:dithered_data}. This target has been observed with a square four-points dither pattern of (3",3").
 
\begin{table*}
\caption{Single epoch - dithered data.}
\begin{center}
\begin{tabular}{lccccccccc}
\hline \hline
Target & Date (UT) & Filter & Texp & \# Im & Ttot & $<$FWHM$>$ & $<$SR$>$ & \# Stars & Pos \\ \hline
NGC2362 & 2011 Dec 15 & H & 15 s & 35 & 525 s & 72 mas & 15\% & 68 & dithered\\ \hline \hline
\end{tabular}
\end{center}
\label{tab:dithered_data}
\end{table*}

\subsection{Data reduction}
GSAOI delivers a 85" x 85" field of view, composed of four arrays with dimensions 41" x 41" and separated by $\sim$3". The GSAOI pixel scale is 20 mas \citep{carrasco2012results}.  All the data set are sky subtracted, and flat fielded. Skies are built from data sets taken before or after the astrometric observations, and either extracted from dithered data, either from dedicated telescope pointing offsets. For some data sets, not enough data is available to build a proper sky, and a sky from a previous night had to be used. We checked that the astrometric performance was not affected by the use of different skies (see Sect. \ref{sec:discussion}). Also, as most of the data has been taken close to Zenith, no corrections were made for differential atmospheric refraction. This is further discussed in Sect. \ref{sec:discussion}.

\subsection{Distortion correction}
\label{ssec:disto}

Frame to frame star positions are impacted by residual distortion over the field. 
In order to remove those distortions, we compensate each frame with a high-order polynomial fit. We used the following definition of polynomials:
\begin{dmath}
x' = c(1) + c(2)*x + c(3)*y + c(4)*x^2 \\ + c(5)*x*y + c(6)*y^2 + ... \\ 
y' \hiderel{=} d(1) + d(2)*y + d(3)*x + d(4)*y^2 \\ + d(5)*y*x + d(6)*x^2 + ... \\
\end{dmath}
Where the $c$ and $d$ coefficients are free parameters and we use the same number of free parameters per axis. 

\begin{figure}
  \caption[] {Astrometric gain after fitting and removing a given number of polynomials transformations (i.e. number of degrees of freedom). The astrometric gain is computed with respect to the performance when only 3 free parameters per axis are used. Black solid line is when the four GSAOI array are considered together. Blue line is when each array is considered independently. }
  \includegraphics[width = 1.0\linewidth]{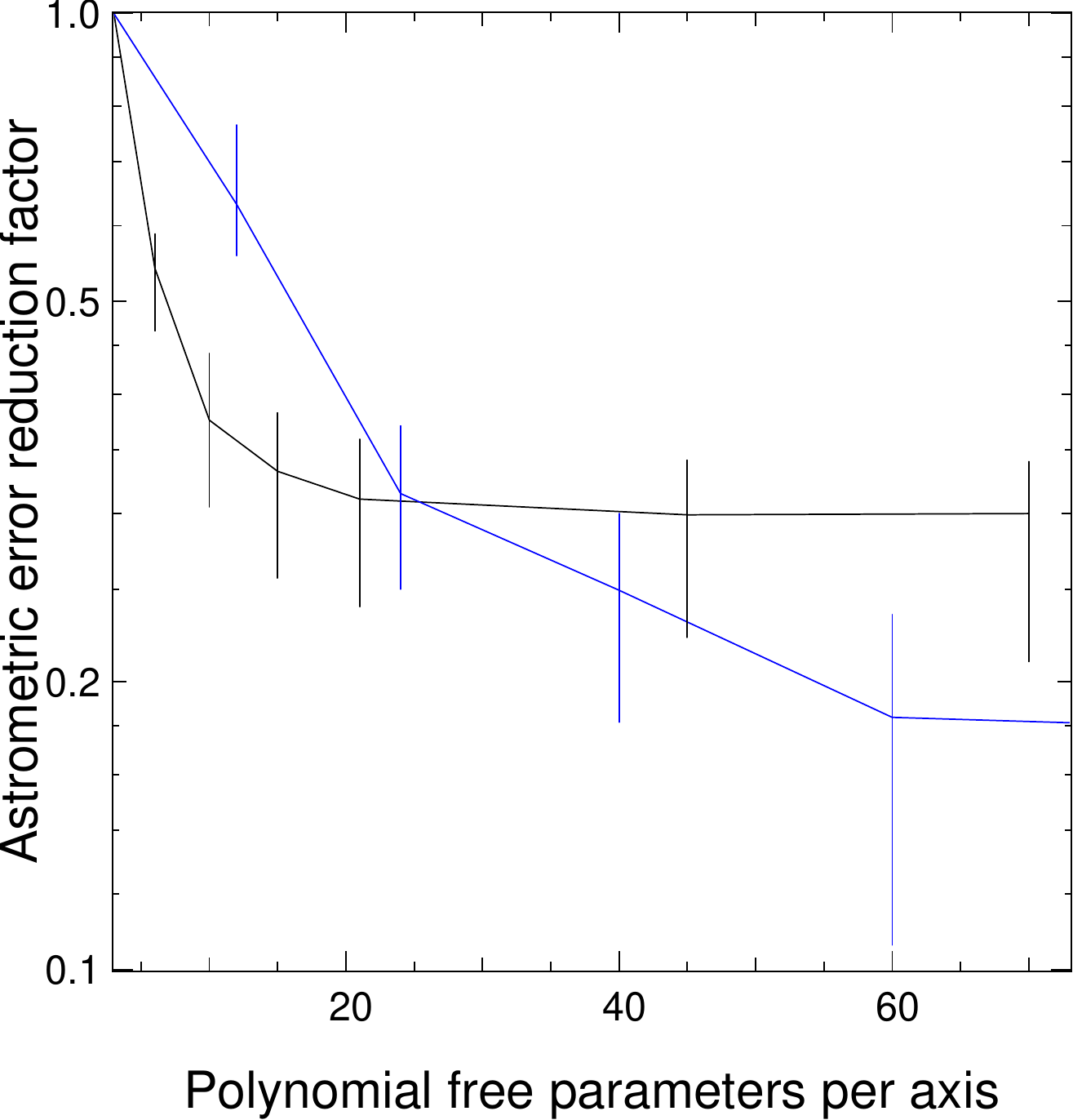}
  \label{fig:poly}
\end{figure}

As a first insight into the nature of the optical distortion present in the images, we have tested the impact of fitting and removing high-order polynomials for the single-epoch, un-dithered data. Figure \ref{fig:poly} shows the residual astrometric error, averaged over all stars and over the full FoV, when an increasing number of polynomials are used. All images are referenced with at least 3 free parameters per axis (6 total), which include Tip, Tilt and rotation. The resulting astrometric error after removing these modes is taken as reference. Then we computed the gain with respect to this baseline, when increasing the number of free parameters. The solid black line shows the astrometric gain when the four GSAOI array are mosaiced together. The error bars show the minimum and maximum astrometric gain for all the images analyzed. The blue solid curve shows the same astrometric gain when each array is treated independently, the number of free parameter reported in Fig. \ref{fig:poly} being the sum for the four arrays. 
Figure \ref{fig:poly} shows that a very signifiant gain ($>$ 80\%) can be reached by compensating for the image distortion, however this compensation requires a fairly large number of degrees of freedom ($>$ 60), hence at least an equivalent number of stars available in each image. This is what set our definition of crowded fields.
We also note that, ideally, for a given number of free parameters, an optimal management of the noise (e.g. \cite{cameron2009}) should give the exact same results if we would treat the full array or each chip independently. We have not implemented such methods for the current analysis, but cutting the array in sub-pieces relaxes the constraint on the noise propagation: for each array, it requires lower order polynomial, which are less sensitive to noise. Instead of using polynomial fit, a better approach could be to describe the distortions based on two-dimensions splines \citep{yelda2010astrometry}. This method seems to be the more robust to noise, and will be discussed in more details in a companion paper (Ammons et al. in prep.). In the following, and unless specified, we will use 15 free parameters per axis applied to each array to compensate for distortions between frames.

\section{Results}
\label{sec:results}

\subsection{Single epoch, un-dithered data}
\label{ssec:singlepoch}

\begin{figure}
  \caption[] {Distribution of the astrometric error over the field for a single 15s exposure image, taken from the December 19th data set. Blue circles show the photon noise limit. The black stars show the location of the three tip-tilt guide stars, the dashed triangle draw the NGS asterism.}
  \includegraphics[width = 1.0\linewidth]{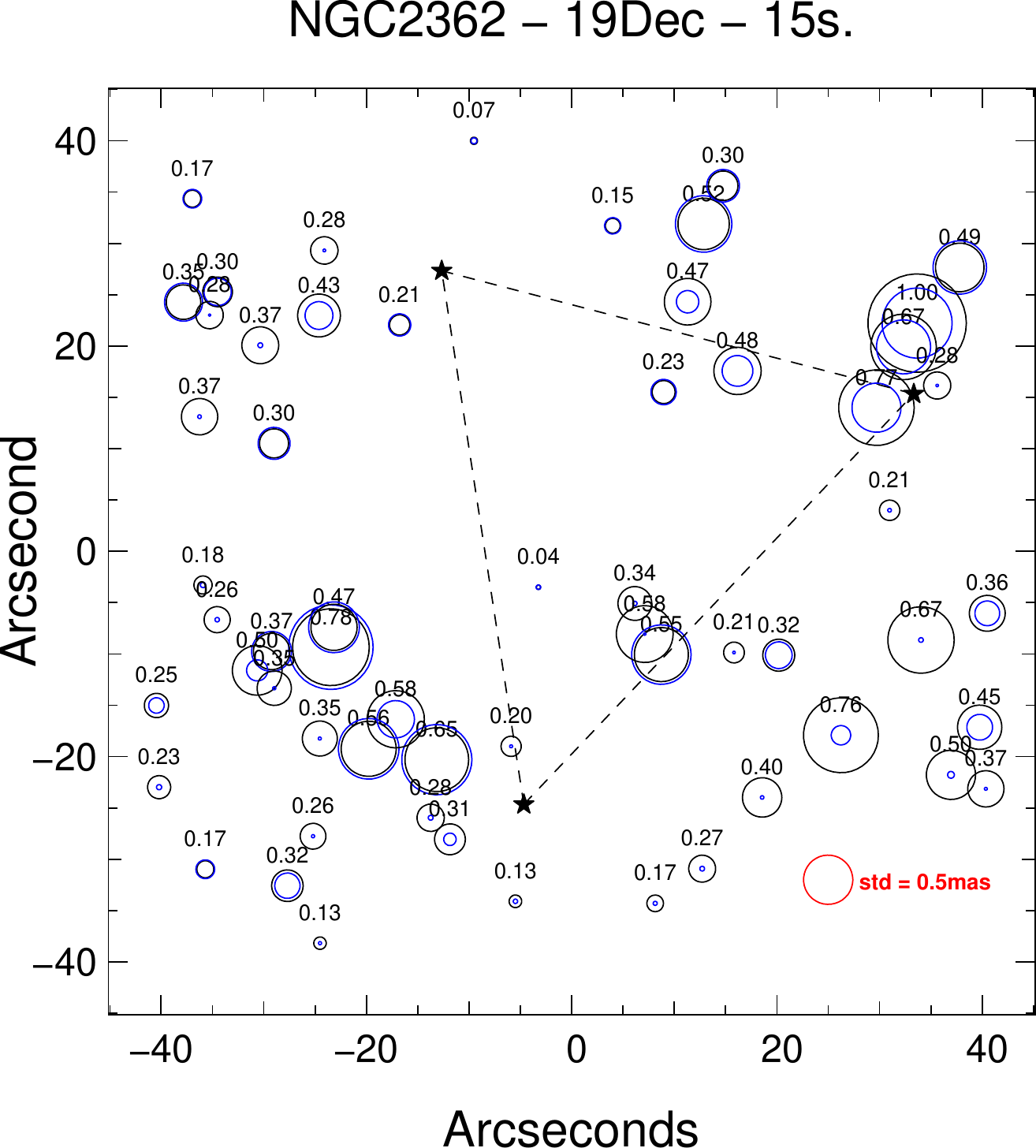}
  \label{fig:ngc2362a}
\end{figure}

We first analyze the data set from December 19th 2011, reported as position 1 in table \ref{tab:singleepoch_data}.
As for the simulations, we first look at the error position over the full images data set (i.e. 27 images in this case), and how these errors are distributed over the field. This is shown in Fig. \ref{fig:ngc2362a}. For reference, the blue circles show the errors due to photon and sky subtraction noise, as derived from the simulations and Fig. \ref{fig:astro_error1}. From Fig. \ref{fig:ngc2362a}, we can see that, although the distribution of errors is quite complex, there is a tendency for it to be larger outside of the Tip-Tilt guide star asterism -marked by the triangle, which is expected as tip-tilt and plate scales are not controlled in this region, and rotation effects are amplified. PSFs variations are also expected to be larger outside the Tip-Tilt guide star asterism, which is impacting the astrometric performance as seen in Sect. \ref{sec:simmul}. We also note that the computed position errors generally agree with the noise estimate, although there is some scatter. Finally, we conclude that positions computed with this method lead to an estimate of the astrometric error of around 0.4 mas.

\begin{figure}
  \caption[] {Astrometric error vs. exposure time for NGC2362 - December 19th 2012 data set.}
  \includegraphics[width = 1.0\linewidth]{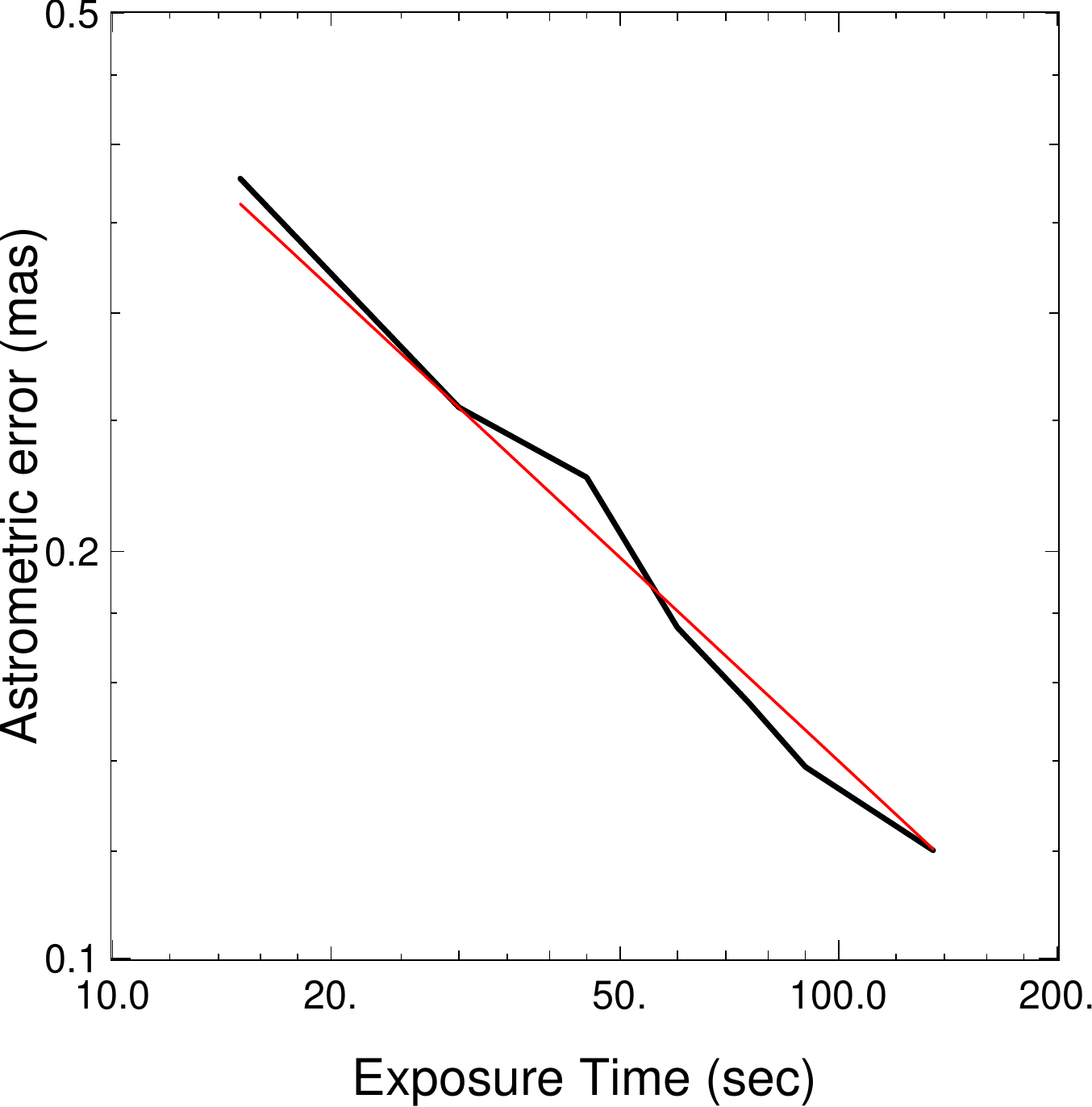}
  \label{fig:ngc2362e}
\end{figure}

\begin{figure}
  \caption[] {Distribution of the astrometric error over the field for the averaged 135 s exposure image, taken from the December 19th data set. Blue circles show the photon noise limit.}
  \includegraphics[width = 1.0\linewidth]{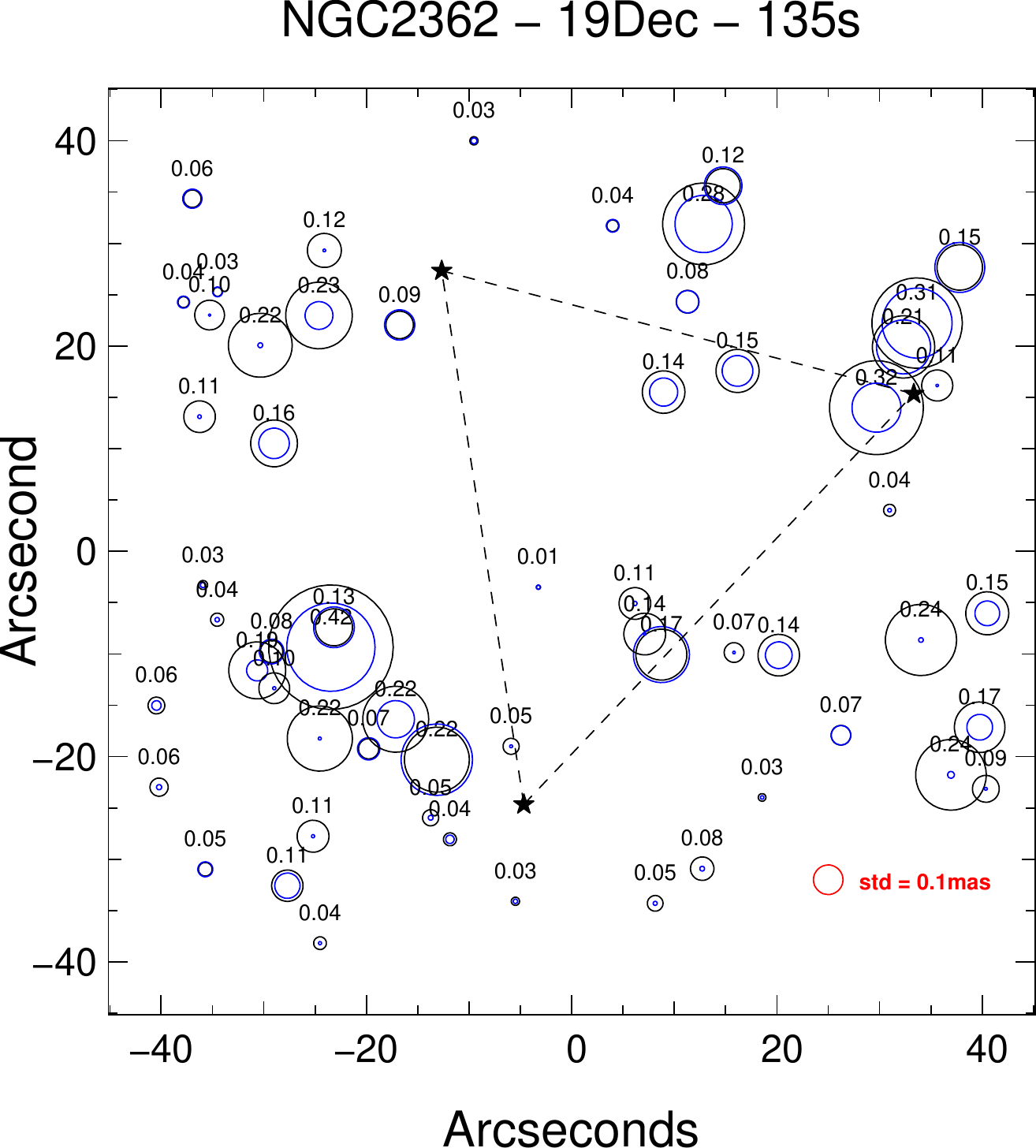}
  \label{fig:ngc2362f}
\end{figure}

In a next step, we explore how the astrometric error scales with exposure time. For this, we combine the mapped images into groups of sub-images. This increases the effective integration time in order to identify any systematic error that does not average out. Results are shown in Fig. \ref{fig:ngc2362e}, the bold black line shows the average of the four arrays, the red solid line is a linear fit (in log-log space). Results from Fig. \ref{fig:ngc2362e} show that the errors are properly scaling with the integration time, and, at least for this data set, no systematic error floor can be detected. 
For the full 135 s. combined data set, the averaged of the four array treated independently gives an astrometric error as low as $\sim$150 $\mu$as. The distribution of the error across the field is shown in Fig. \ref{fig:ngc2362f}. Note that the scale is different from the one in Fig. \ref{fig:ngc2362a}.
 
\begin{figure}
  \caption[] {Pairwise astrometric error as a function of the distance between stars for the NGC2362 December 19th data set. Solid lines are the median value per 10\arcsec separation bin. Fifteen degrees of freedom per array are used to register the frames.}
  \includegraphics[width = 1.0\linewidth]{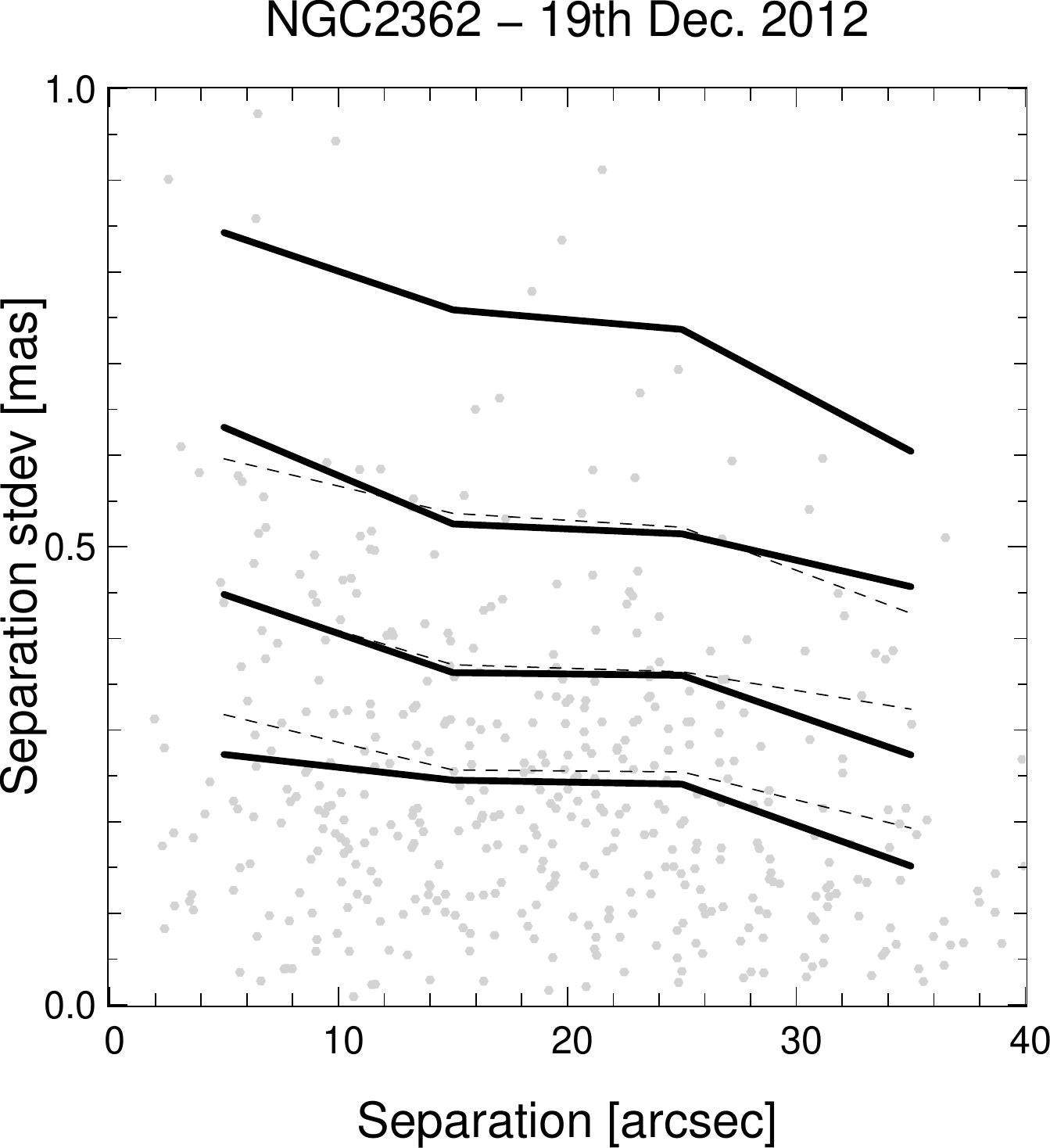}
  \label{fig:ngc2362g}
\end{figure}

\begin{figure}
  \caption[] {Same as Fig. \ref{fig:ngc2362g}, but only three degrees of freedom per array are allowed to register the frames.}
  \includegraphics[width = 1.0\linewidth]{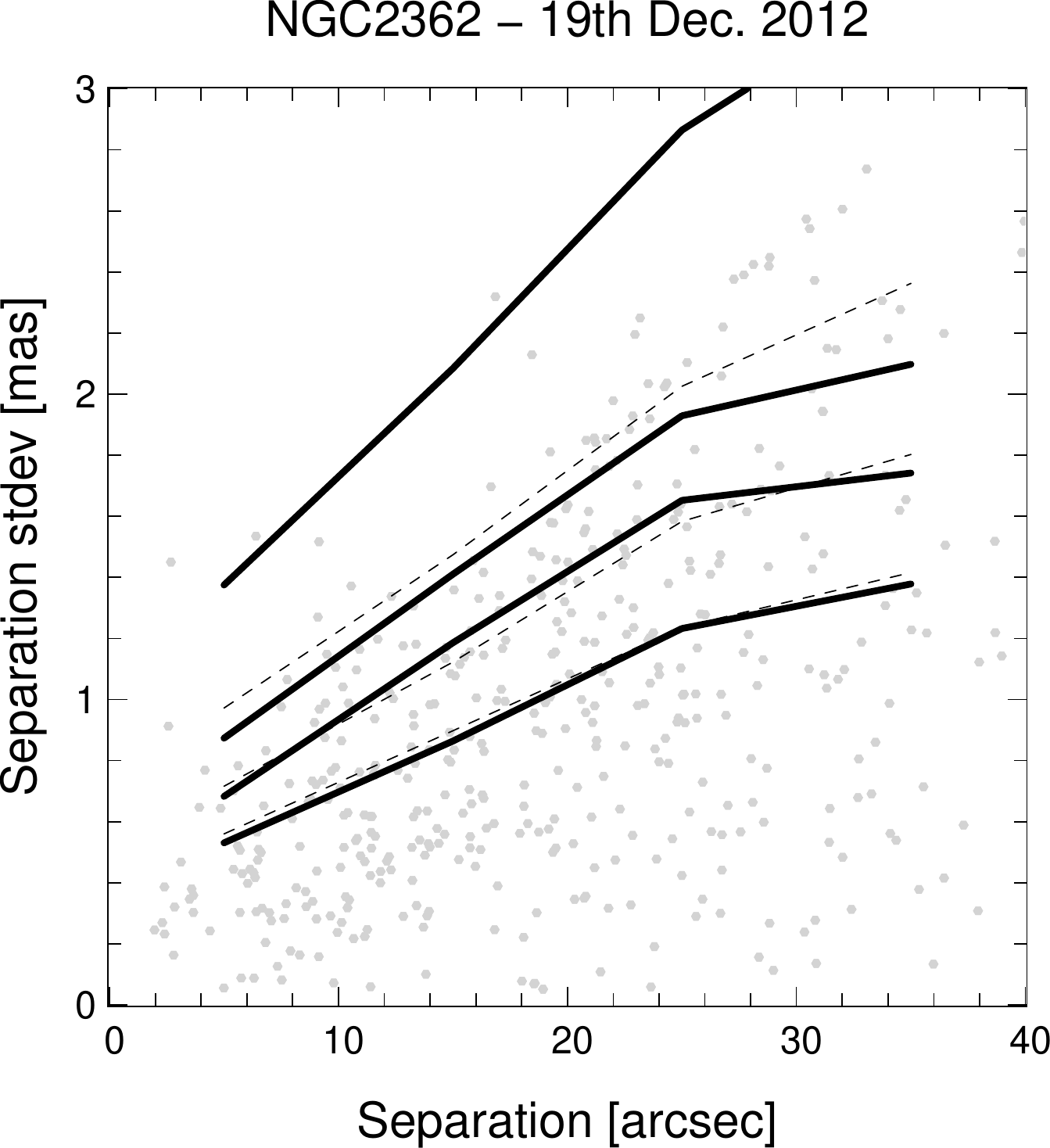}
  \label{fig:ngc2362h}
\end{figure}

In Fig. \ref{fig:ngc2362g}, we show how the pairwise astrometric error is distributed with the distance between stars. As for Fig. \ref{fig:astro_error4}, the solid lines show the median of the errors per 10\arcsec bin, for different numbers of binned images or, equivalently, different exposure times. The top curve shows the single-exposure case (no binning of images). The bottom curve corresponds to a 135 s integration time (i.e. 9 images binned). The dashed lines show the errors, scaled by the square root of the integration time, and it follows fairly well the measured errors. This is fully consistent with the results of Fig. \ref{fig:ngc2362e}. 
Fig. \ref{fig:ngc2362h} shows the same as Fig. \ref{fig:ngc2362g}, but when only three degree of freedom per array are used (tip, tilt and rotation). In such a case, field distortions have note been properly removed, and the pairwise error increases as the distance between the star increases. 

\begin{figure}
  \caption[] {Astrometric error vs. exposure time for all single epoch, un-dithered data sets (see table \ref{tab:singleepoch_data}). Square symbols are for the NGC2362 data, star symbols for the NGC1851 data, and triangles for the LMC data. }
  \includegraphics[width = 1.0\linewidth]{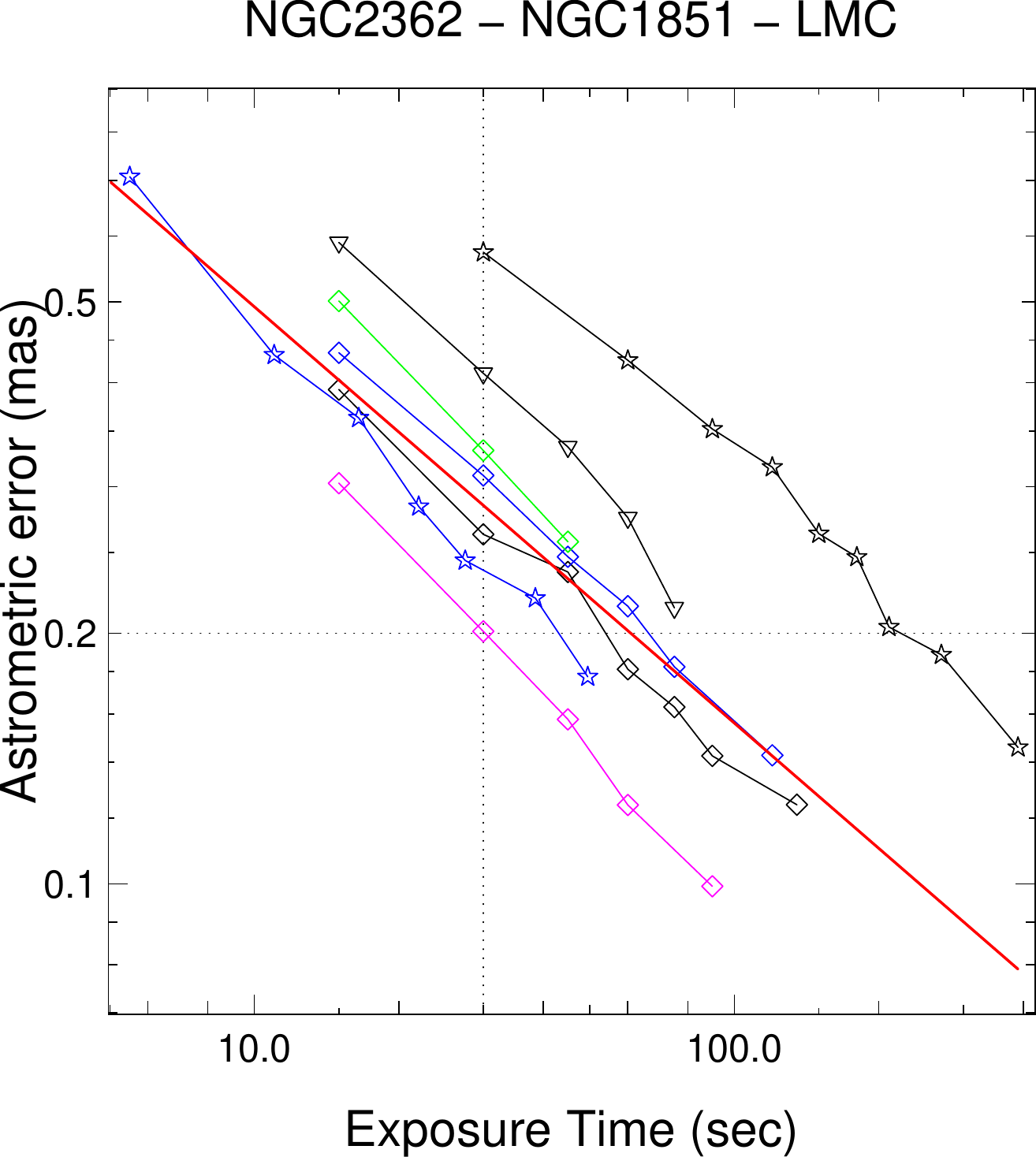}
  \label{fig:ngc2362i}
\end{figure}

\begin{figure}
  \caption[] {Astrometric error as a function of the field-averaged SR, measured in H-band.}
  \includegraphics[width = 1.0\linewidth]{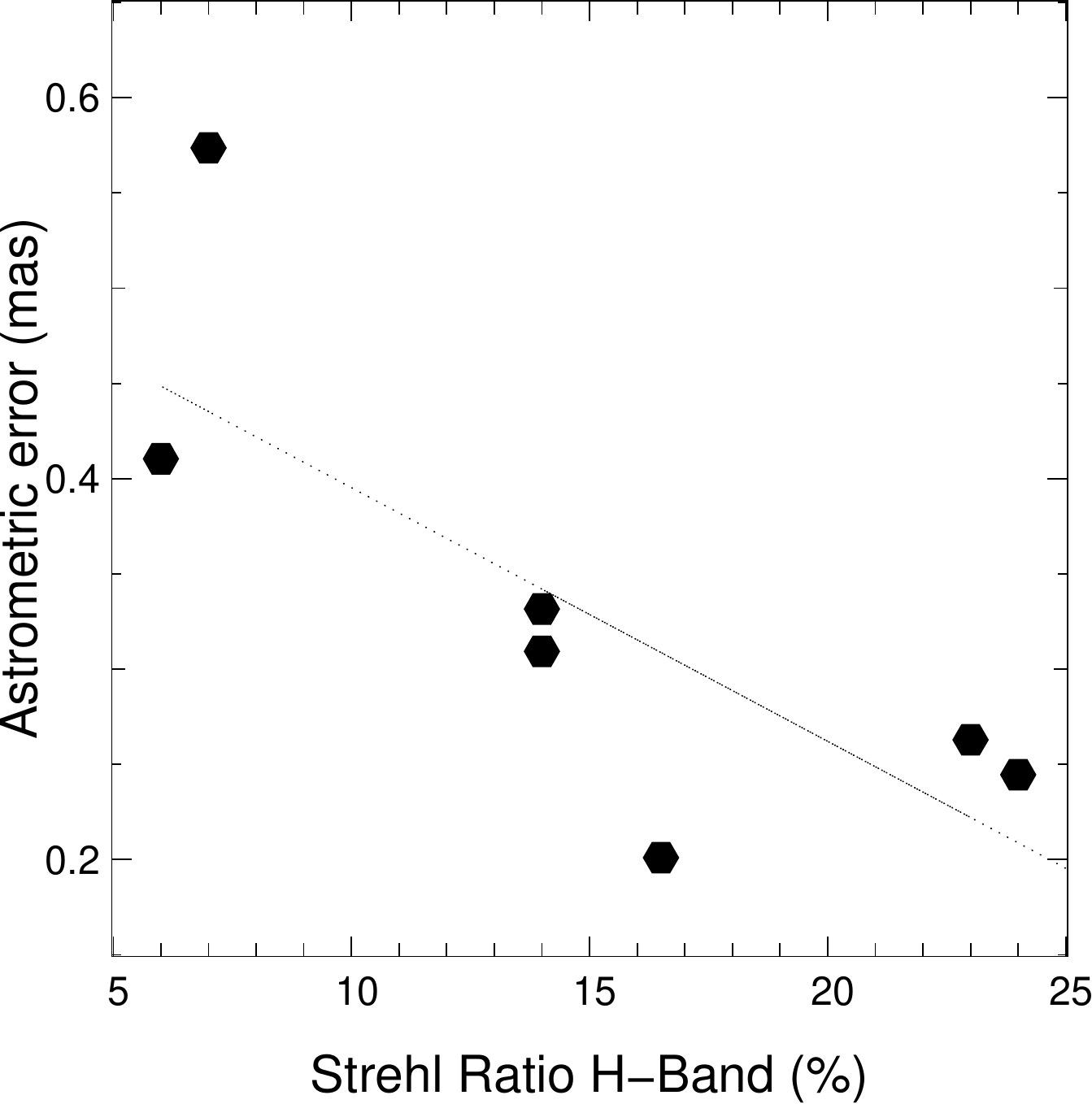}
  \label{fig:ngc1851aa}
\end{figure}

We have analyzed all the other data set presented in Tab. \ref{tab:singleepoch_data} in a similar manner. Results are presented in Fig. \ref{fig:ngc2362i}. Square symbols are for the NGC2362 data, star symbols for the NGC1851 data, and triangles for the LMC data. All the data set follow a linear decrease with the square root of the integration time, and no systematic are detected here. The differences in absolute performance is  explained by differences in AO performance. Indeed, if we report the astrometric error at a given exposure time (for instance 30 s.) versus the averaged SR of the images, we get the result presented in Fig. \ref{fig:ngc1851aa}: the astrometric performance is well correlated with the SR.\\

\begin{figure}
  \caption[] {Astrometric error as a function of the star magnitudes for NGC1851 - Nov. 05th 2012. The solid lines show the limits imposed by noise and PSF variations.}
  \includegraphics[width = 1.0\linewidth]{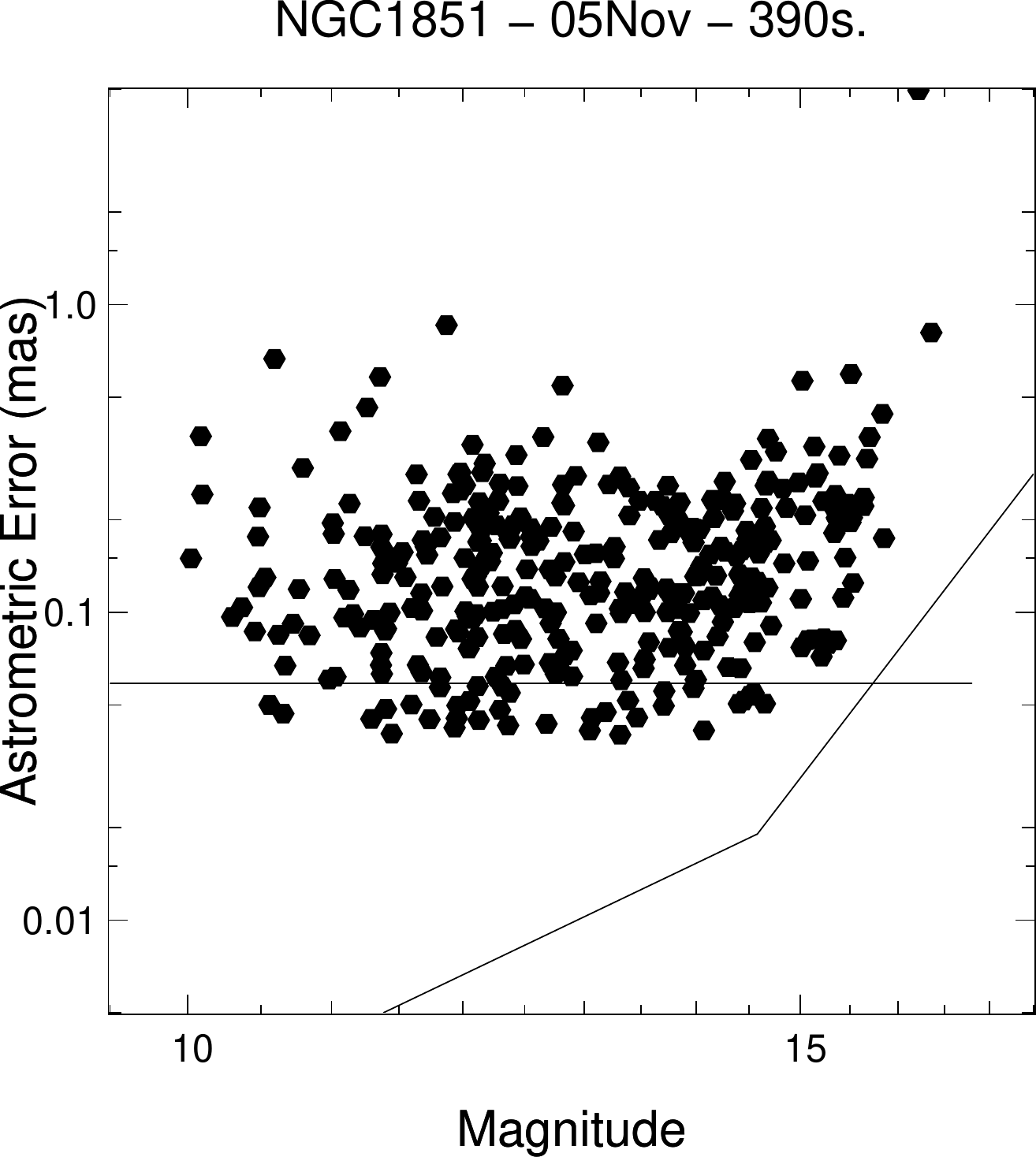}
  \label{fig:ngc1851a}
\end{figure}

With the NGC1851 data set, we explored how the astrometric error scales with the star magnitudes, and we compared our results with the one obtained from the simulations in Sect. \ref{sec:simmul}. This is presented in Fig. \ref{fig:ngc1851a} were the points are all the stars selected (621 stars), and the solid lines show the fundamental limits imposed by noise, and the derived plateau due to PSF variations over the field (horizontal line). Both errors have been scaled to the integration time obtained on NGC1851 (390 s). A good agreement between the simulations and the measurements is seen: most of the data points are close to the theoretical limits. This also means that PSF variations could explain the performance we observe. \\

\begin{figure}
  \caption[] {NGC1851 - The three circles show the regions used in the impact of crowding analysis. The two right-hand side regions are averaged and reported as ``outer" in Tab. \ref{tab:ngc1851_crowd}. The left-hand side region, which encompass the cluster center, is reported as ``inner" in Tab. \ref{tab:ngc1851_crowd}.}
  \includegraphics[width = 1.0\linewidth]{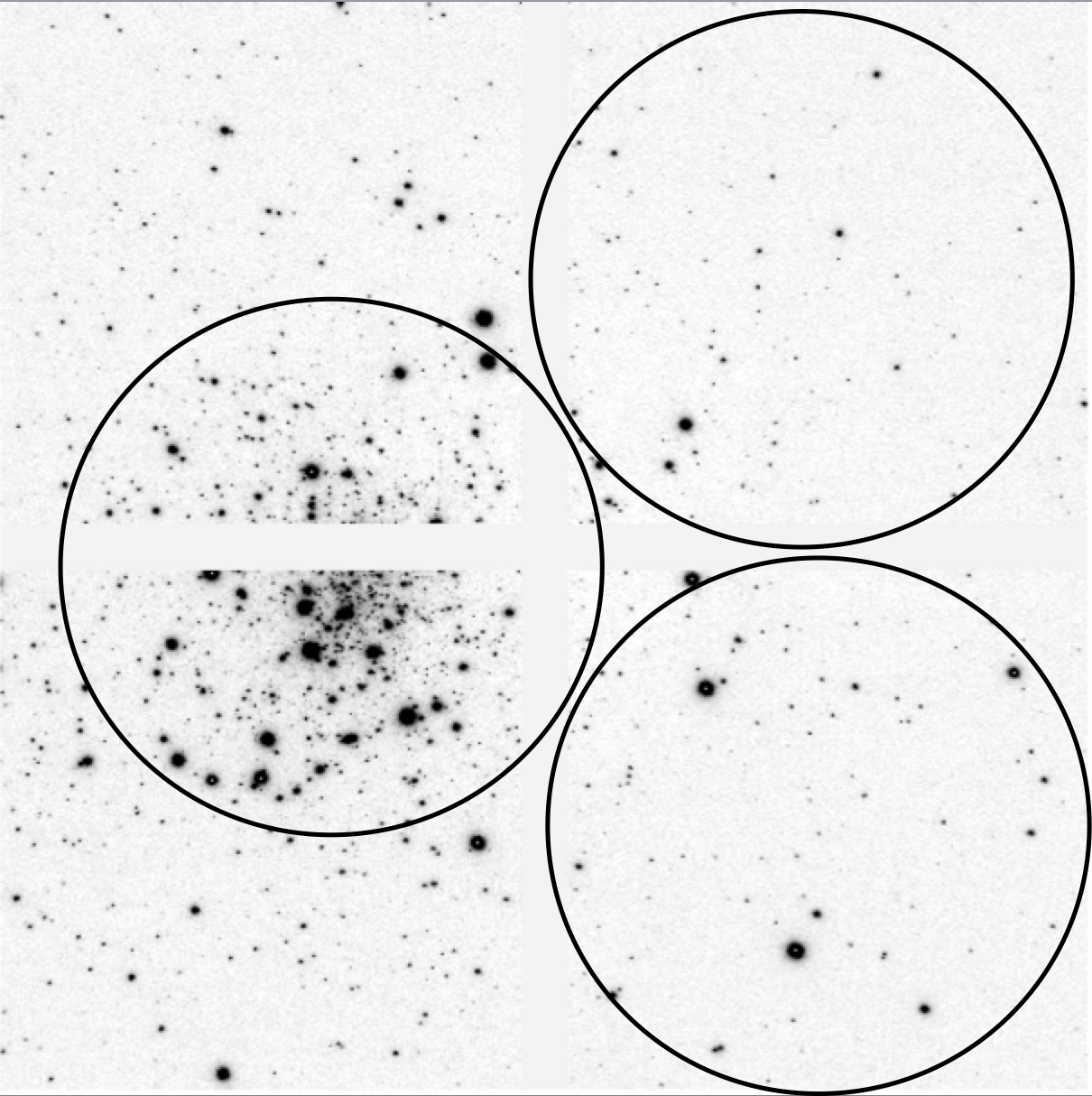}
  \label{fig:ngc1851crowd}
\end{figure}

Finally, for NGC1851, we looked at the astrometric error when considering sub-fields with different crowding level. In Fig. \ref{fig:ngc1851crowd} we show the three sub-fields considered: one at the center of the cluster, and two fields located outside the cluster center. The goal here is to test the impact of crowding on astrometric performance, by comparing the center field performance, with the two outer regions taken as reference. For this, we use the three NGC1851 data sets presented in Table \ref{tab:multiepoch_data}. Each region is analyzed independently, and mapped with 15 degrees of freedom. Results are presented in Table \ref{tab:ngc1851_crowd}, and show that the error is higher in the central part of the cluster than in the outskirt, most probably due to the crowding effect. Hence, the use of advanced techniques to explicitly take the crowding into account such as the one presented in \cite{schodel2010} should be considered for the most crowded fields.

\begin{table*}
\caption{Impact of crowding on NGC1851.}
\begin{center}
\begin{tabular}{cc|cc|cc}
\hline \hline
\multicolumn{2}{c|}{30th Dec. 2012} & \multicolumn{2}{c|}{31th Dec. 2012} & \multicolumn{2}{c}{28th Jan. 2013}\\ \hline
inner & outer & inner & outer & inner & outer \\
0.87 mas & 0.75 mas & 0.90 mas & 0.71 mas & 0.62 mas & 0.54 mas \\ \hline \hline
\end{tabular}
\end{center}
\label{tab:ngc1851_crowd}
\end{table*}

\subsection{Single epoch dithered data}
\label{ssec:dithered}
\begin{figure}
  \caption[] {Comparison of the astrometric error when using un-dithered and dithered data. Black solid lines are for un-dithered data. Magenta curves are for dithered data. The two top curves are when 6 degrees of freedom per array are used to map the coordinated. The bottom two curves are when 15 transformation parameters are used.}
  \includegraphics[width = 1.0\linewidth]{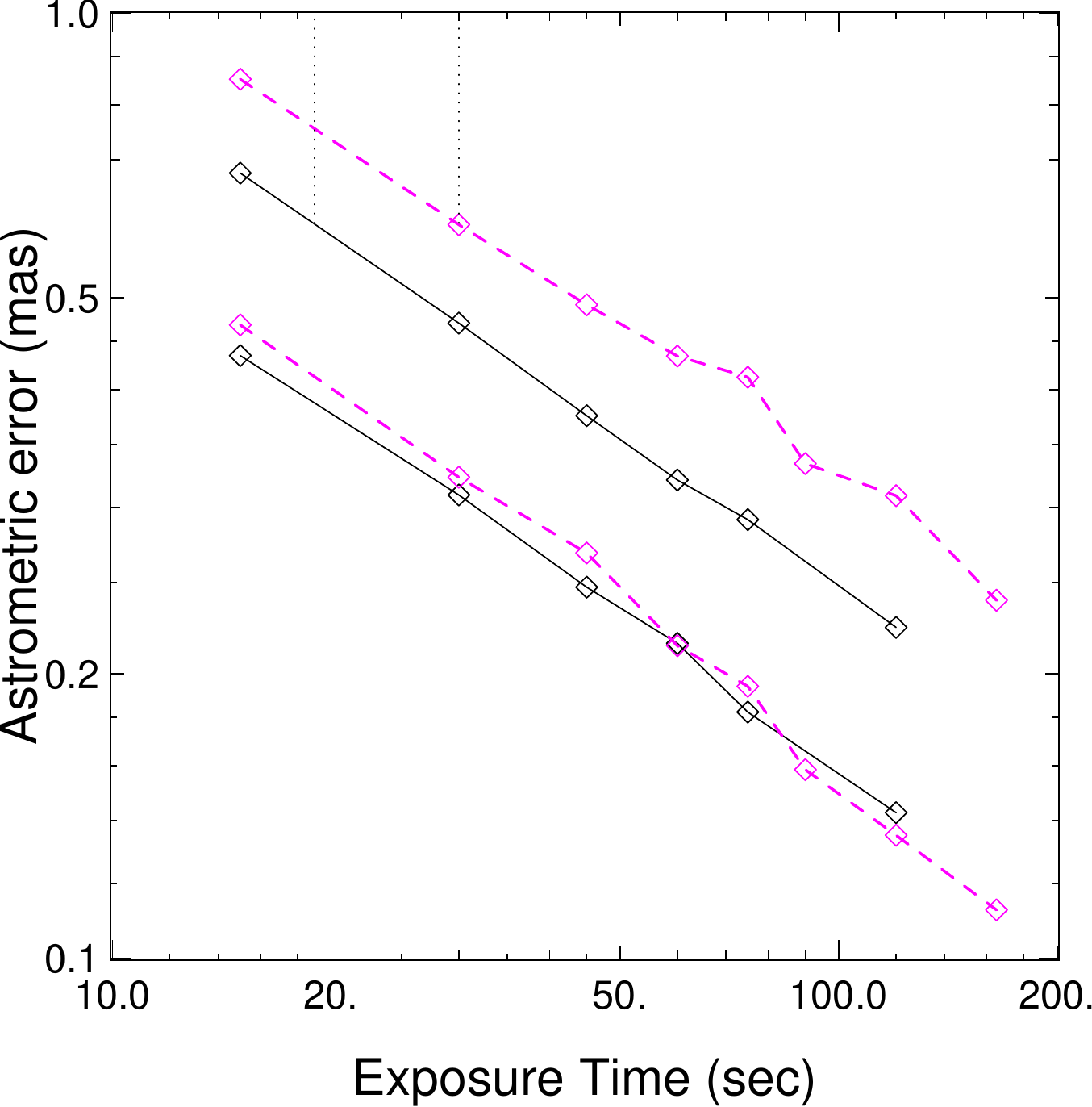}
  \label{fig:ngc2362-dither}
\end{figure}

We have seen so far that the astrometric performance on single epoch, un-dithered data could be as good as $\sim$ 150 $\mu$mas, if enough stars are available in the field to filter high-order distortions present in the images. These high-order distortions will affect the performance when the image is dithered on the detector, as each star will see a different distortion pattern. If these distortions are of high-order, it might even be impossible to remove them all. In this section, we explore the astrometric performance when using dithered data. For this, we use the data presented in Table \ref{tab:dithered_data}. We have 35 images that have been taken with a 4-points square dither of (3", 3"). Results are analyzed as the astrometric error versus integration time, and are presented in Fig. \ref{fig:ngc2362-dither}. The two black lines show the error when 6 (respectively 15) degrees of freedom per chip are used to map the images, for un-dithered data. The magenta lines show the same, but for dithered data. Dithering affects the astrometric performance when only 6 degrees of freedom per chip, say for low-density fields. In this case, one would require almost twice the integration time when dithering than without dithering. For high-density fields, images can be dithered with almost no penalty if at least 15 degrees of freedom per detector are used in the image transformations. 

\subsection{Multi epoch data}
\label{ssec:multipoch}

\begin{figure}
  \caption[] {Difference between star positions measured on two consecutive nights with individual frames transformed into a common coordinate system with three degrees of freedom per array. The RMS error of the positional differences is 2.6 mas.}
  \includegraphics[width = 1.0\linewidth]{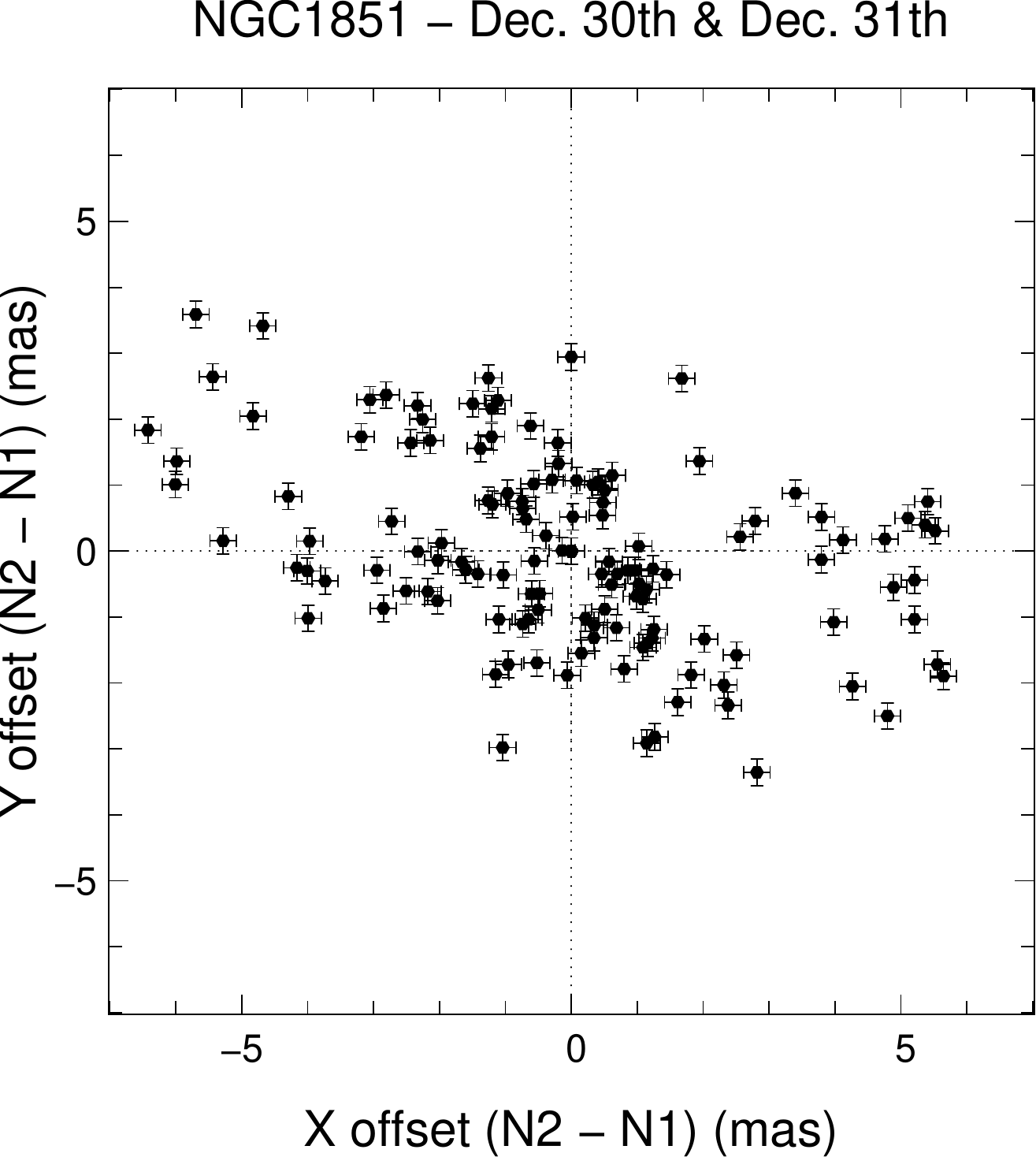}
  \label{fig:ngc1851-multiepoch2}
\end{figure}

\begin{figure}
  \caption[] {Same as Fig. \ref{fig:ngc1851-multiepoch2}, but 15 degrees of freedom per array are used to map the frames. The RMS error of the positional differences is 0.55 mas.}
  \includegraphics[width = 1.0\linewidth]{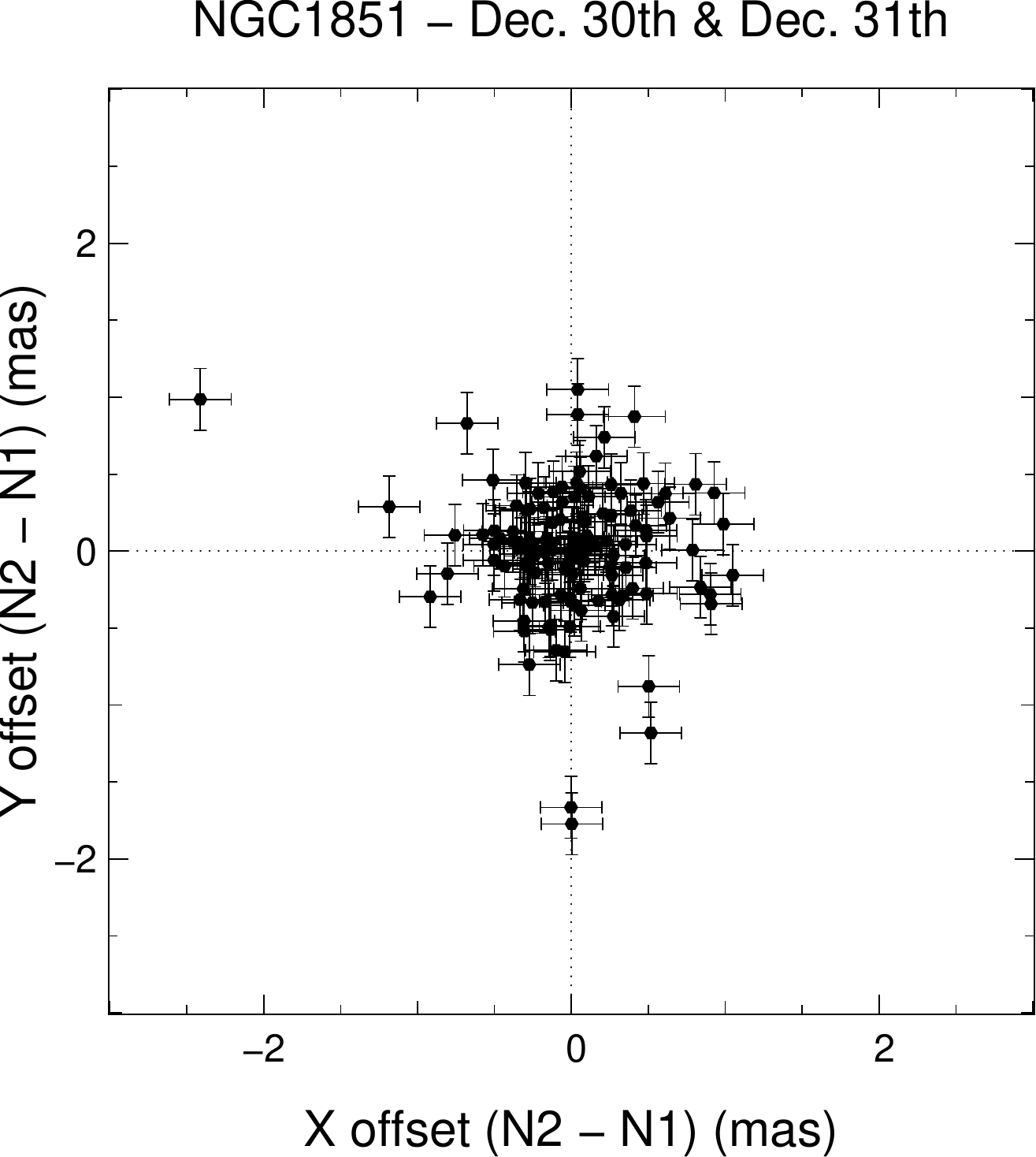}
  \label{fig:ngc1851-multiepoch1}
\end{figure}

Astrometric programs typically need to  reproducing a given observation over a long period of time, to detect proper motions or parallaxes of the sources. Hence, the multi-epoch astrometric errors need to be properly understood. In this section we make use of data that has been observed over different period of time, from 1 night to more than 1 year, to evaluate potential systematics errors. For each data set presented in Table \ref{tab:multiepoch_data}, we do the following. First, a master-coordinate reference frame is built by averaging starlists over all epochs after applying distortions corrections. Then, individual starlists are transformed to the master-coordinate reference frame and the starlists in each epoch are averaged to make one starlist per epoch. Finally, we compare different epochs by computing the difference between star positions in two consecutive epochs. Results for NGC1851 from two consecutive nights are shown in Fig. \ref{fig:ngc1851-multiepoch2} and Fig. \ref{fig:ngc1851-multiepoch1} for transformations using 3 and 15 degrees of freedom per detector, respectively. The error bars are from the single-epoch analysis and are typially 0.2 mas. The multi-epoch astrometry is less accurate than predicted based on the single-night precision with a residual RMS error of 2.6 mas for three degrees of freedom per array, and 0.55 mas for 15 degrees of freedom. The high-order transformation removes most of the time variable distortion; however, some spatial correlations remain, as evidence by the asymmetry seen in Fig. \ref{fig:ngc1851-multiepoch1}. This suggests that even higher-order residual distortions are still present, introducing systematic errors. To quantify this systematic error term, we assume that the resulting scatter (RMS error) is the sum of a random component, taken as the single-epoch error, and a systematic component. Assuming a $\sim$0.2 mas of random error per epoch, the remaining systematic error would be $\sim$0.45 mas. Restricting the analysis to the brightest 50\%  of stars, and a FoV of 30\arcsec x 30\arcsec, this systematic noise floor is reduced to 0.3 mas. 

\begin{figure}
  \caption[] {Multi-epoch astrometric error as a function of the total number of degrees of freedom used to map the frames together.}
  \includegraphics[width = 1.0\linewidth]{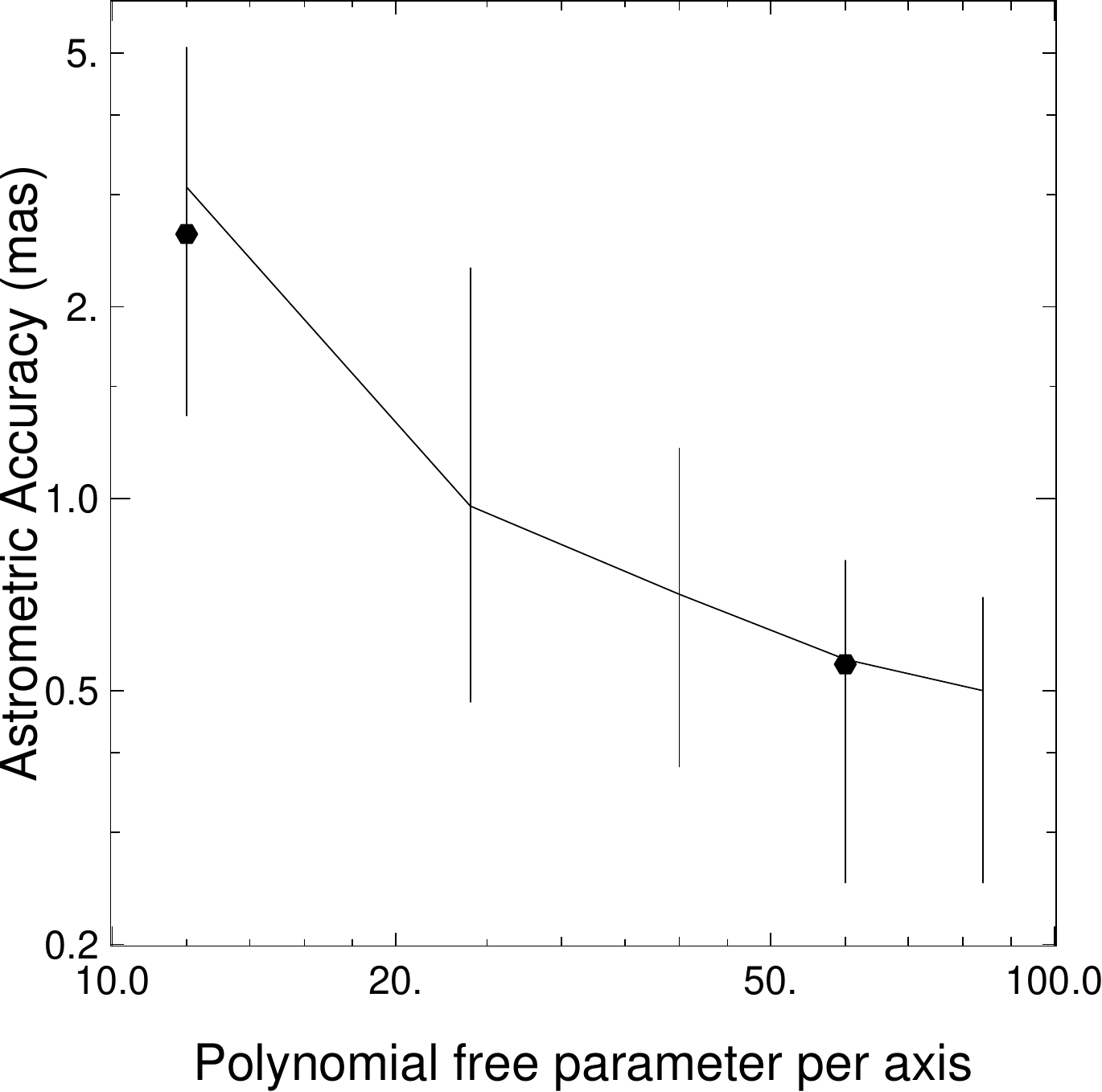}
  \label{fig:multiepoch}
\end{figure}

In Fig. \ref{fig:multiepoch} we analyze, for all the data presented in Table \ref{tab:multiepoch_data} this time, how the multi-epoch error scales with the total number of degree of freedoms used to map the frames together. Errors bars represent the minimum and maximum errors obtained over all the images for each case. Single epoch error has not been quadratically subtracted, and the degrees of freedom quoted are for the full array (i.e. four times the number of degrees of freedom per quadrant). For reference, results obtained in Fig. \ref{fig:ngc1851-multiepoch2} and \ref{fig:ngc1851-multiepoch1} are over-plotted as black dots in Fig. \ref{fig:multiepoch}. The trend shows that more degrees of freedom reduces the systematic errors observed over different epochs. However, large residuals are still present, which again suggests the remaining distortions are of high-order nature. These results, as well as alternative methods to map and remove the distortions between epochs are discussed in a companion paper (Ammons et al. in prep.). 

\begin{figure}
  \caption[] {Vector differences when comparing 2 epochs. Vectors have been amplified by 2500 to be in arc-seconds.}
  \includegraphics[width = 1.0\linewidth]{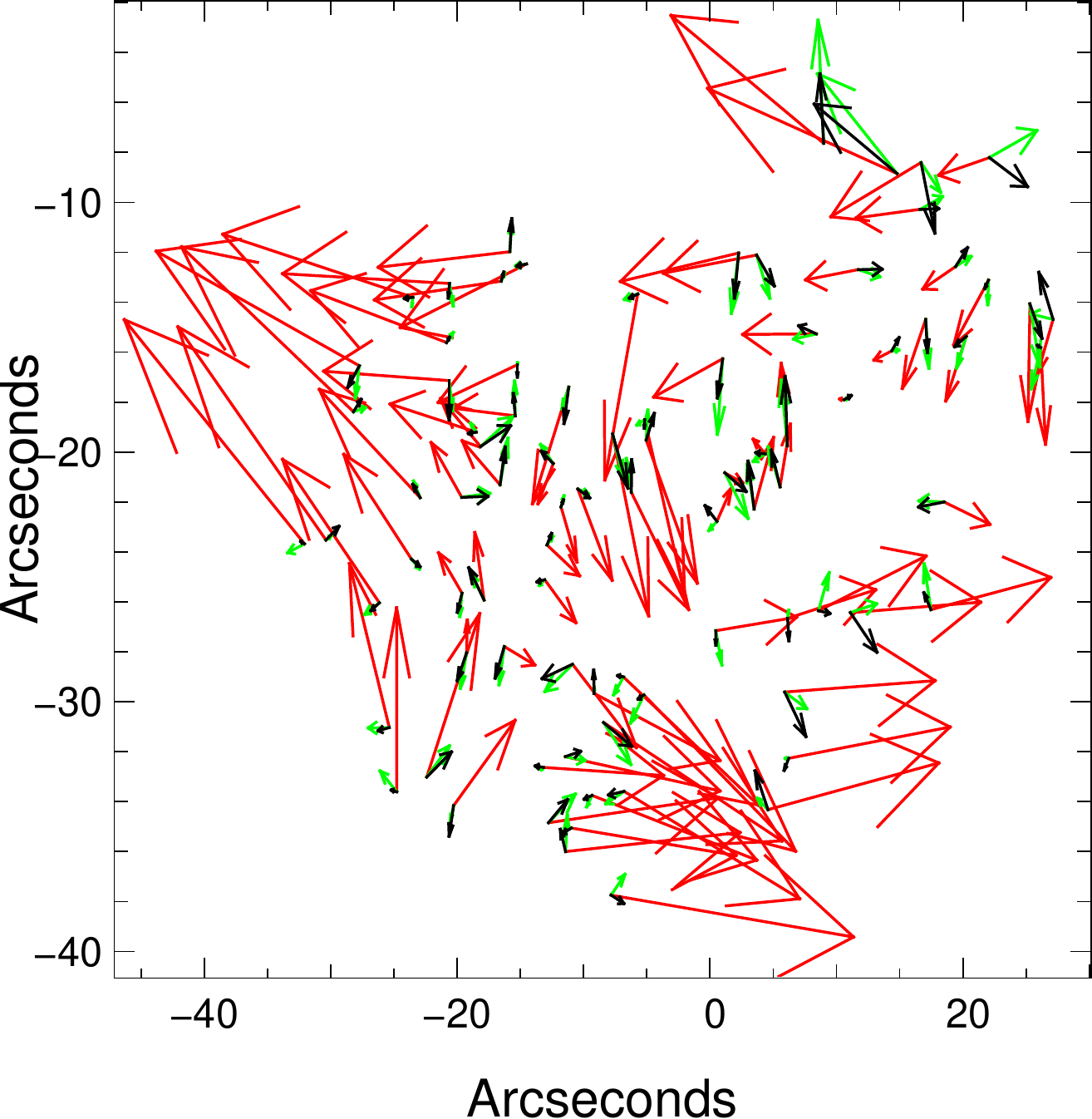}
  \label{fig:ngc1851-vectors}
\end{figure}

Finally, in Fig. \ref{fig:ngc1851-vectors}, we display the vector differences between the stars’ positions in the averaged image of NGC1851 of the 30th of December, as compared to those of 31th of December. In red are the vector differences when only three degrees of freedom per array are used, in green is for 10 degrees of freedom and in black is for 15. Fig. \ref{fig:ngc1851-vectors} illustrates the nature of the multi-epoch distortions: mostly low orders as can be seen from the red arrows. However, after fitting and removing low-order terms, large high-order residuals remains, as can be seen from the green and black arrows.

One possible method to improve the multi-epoch performance is to use an absolute reference grid such as background galaxies. 
However, background galaxies are faint and may require stacking many images in a single epoch to get a large sample of reference sources. We tested this scenario by decreasing the number of free parameters used to register a single epoch, but using high-order transformations to register across multiple epochs. We found similar performance when using 15 parameters for the single-epoch registration and 3 for the multi-epoch registration, as when we used 3 for the inter-epoch registration, but 15 for the multi-epoch mapping. Hence, it seems that the total number of degrees of freedom available is the important factor, rather than how they are distributed. We note that using galaxies as reference may be less accurate than stars as there are extended objects \citep{trippe2010}. As a final remark, background galaxies would only appear on long-exposure images, hence single images should be taken with as small or no dithers in order to get the best single-epoch combined image, as pointed out in Sect. \ref{ssec:refframes}.

\subsection{GLAO vs. MCAO}

\begin{figure}
  \caption[] {Average excess of astrometric error between images obtained in GLAO and in MCAO, as a function of the number of degrees of freedom used to map the frames together.}
  \includegraphics[width = 1.0\linewidth]{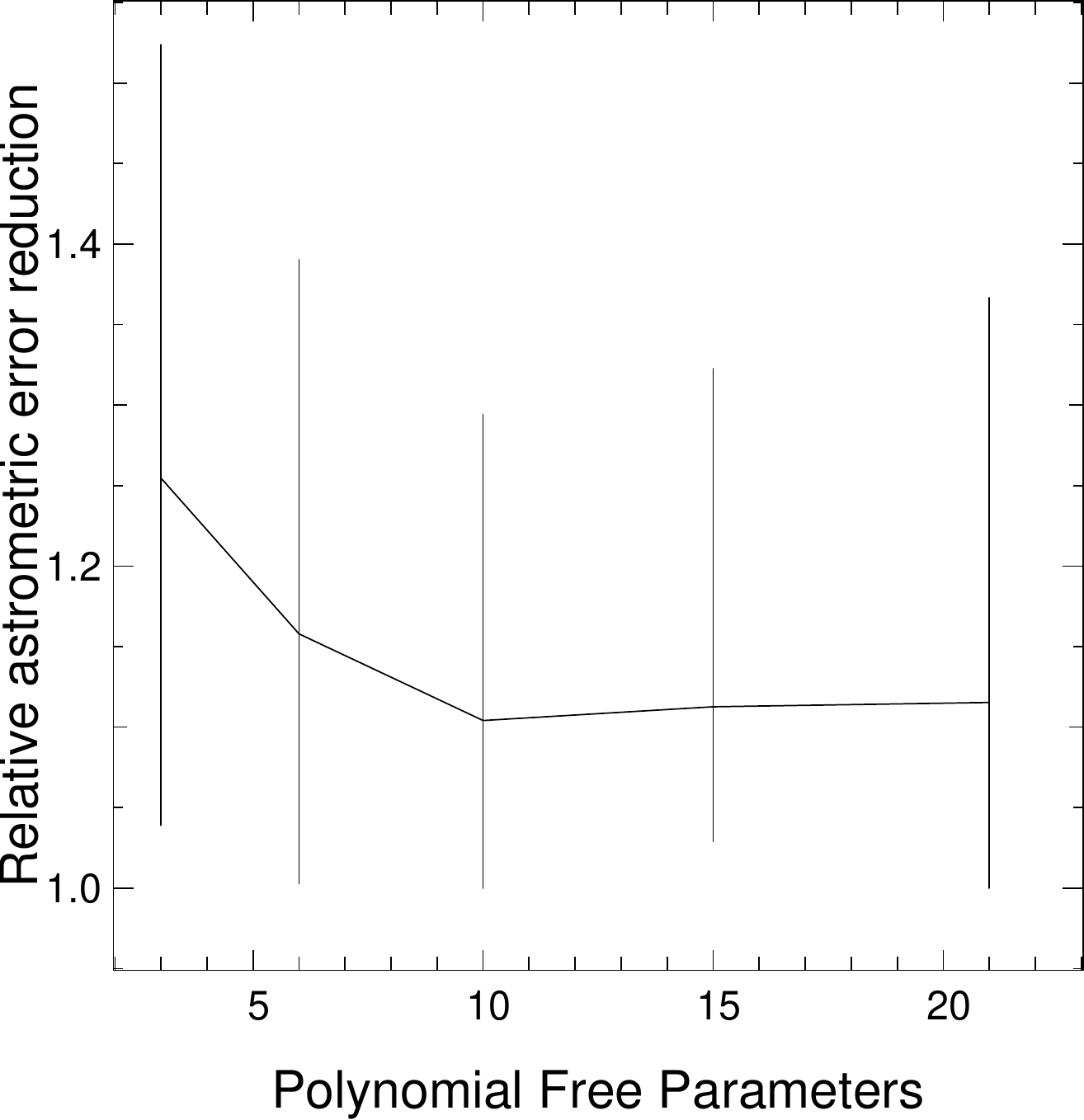}
  \label{fig:glao}
\end{figure}

Over the different commissioning runs, we gathered data taken in MCAO mode (2 DMs) and in GLAO mode (only the ground DM is used). In particular, we found four data sets for which interleaved MCAO-GLAO observations were made. For each data set, 6 images of GLAO and 6 of MCAO are available, interleaved every two images. 
Results are presented in Fig. \ref{fig:glao}. The solid line shows the average excess of astrometric error between the GLAO and the MCAO images, for all the data sets, and for an increasing number of degrees of freedom used to map the images together. For each data set, we took the MCAO astrometric performance as a reference: we divided the GLAO astrometric error by the MCAO astrometric error. The error bars show the minimum and maximum deviation obtained for each case. As state above, the sample is limited in size, and may suffer for some statistical bias, even so it seems that for a low number of degrees of freedom used to map the images together, the GLAO performance is lower than expected just from the difference in AO correction performance between MCAO and GLAO. Indeed, for this specific data set, the GLAO FWHM was $\sim$ 15\% larger in average than the MCAO FWHM. With the altitude DM, MCAO potentially compensates for atmospheric distortion modes, which improves the astrometric performance. With a single DM, conjugated to the pupil, a GLAO system cannot dynamically compensate for such atmospheric distortions. This gain however diminish when higher-order transformation can be used to register the GLAO images.

\section{Discussion}
\label{sec:discussion}

\subsection{Origin of the multi-epoch distortion error}

Over all the sources of error affecting the GeMS/GSAOI astrometric performance, the multi-epoch residual distortion is the main one in the error budget. The origin of the distortion drift has not been clearly identified. It might be due to changes in the gravity vector (the AO bench is mounted on the Gemini Cassegrain focus) or in the environmental parameters (temperature, humidity). The fact that single epoch, un-dithered data also show a dramatic improvement from high-order polynomial transformations suggests that there is a time-variable component to the distortion, as those data sets should not be affected by static optical distortions. The likely source of the time-variable distortion is gravity induced flexure, as this is the environmental parameter changing the most quickly, however only a full correlation of the distortion coefficients with all the environmental parameters is needed to clearly identify the source. Over a single epoch and between epochs, the elevation angle changed  by $\sim$10$^{\circ}$, which may be sufficient to introduce such distortions. For GeMS, the amplitude of the static distortions is estimated to be as large as few arcseconds. Hence even a small drift of the beam will have an impact on the final astrometric performance. 
Another issue with GeMS/GSAOI is that the instrument and the AO bench are regularly removed from the telescope to leave the Cassegrain observing ports free for other instruments. Typically, GSAOI is removed every couple months, and Canopus (the AO bench of GeMS) is removed once a year. Maintenance work and re-installation of these components, even done with particular care, cannot be perfectly reproducible, which might introduce part of the systematic evidenced in this paper.

In crowded fields like the Galactic Center and clusters, the large numbers of stars could be enough to fit high-order polynomials to remove changing distortion. For sparse-field applications, such as using high-precision astrometry on nearby stars to measure masses of orbiting exoplanets, the number of stars in the field is generally not sufficient for this, and alternative methods will be discussed in a companion paper (Ammons et al. in prep.). 


We have looked for other potential sources of errors such as Chromatic Differential Atmospheric Refraction (CDAR), but did not find any obvious correlation between astrometric errors and colors and elevation. All of our data set has been taken at high elevation ($>$70$^{\circ}$), hence CDAR is a second order error term. We have also looked for the variation in centroids due to changes in the flat field from epoch to epoch. For this test, we scanned a simulated 100 mas FWHM Gaussian around the detector, comparing the centroids using a December 2012 flat field (twilight) and a January 2013 dome flat. From that test, we estimate that this error does not contribute for more than 0.1\% of a pixel, which is far below the level of errors measured in the data.

\subsection{Referencing dithered frames}
\label{ssec:refframes}

Distortion also affects the Tip-Tilt WFS focal plane. 
The TT WFS assembly moves as a rigid body. Thus after a dither, the TTWFS are not matched with the TT guide star positions. These static position errors will eventually be compensated by the TT, plate scale and rotation offload to the Cassegrain rotator, inducing offsets, scalings and rotation in the output GSAOI field. The current mitigation plan involves finding an astrometric solution in the dithered images themselves, and compensate for these drifts in the post-processing data reduction.There are only six parameters to determine (2 offsets, 3 plate scale modes and a rotation), so this should generally be doable with at least 3 stars. If less than three stars are detected on each single frame, then referencing the dithered images will be a problem. 
Note also that the science image distortions induced by the TTWFS focal plane distortions depend not only on the constellation position, but obviously also on the constellation itself, so there is no way to easily calibrate on one object and apply for the next object. Solving this problem entirely would involve calibrating the full TTWFS focal plane distortion field. This should be part of a future upgrade of the GeMS NGSWFS \citep{Rigaut2014ngs2}.

\subsection{Performance comparison with other facilities}

How GeMS performs in terms of astrometric performance compared to other facilities has been addressed in \cite{Jessica2014spie}, where the authors compare the performance of GeMS/GSAOI, Keck NIRC2, and the HST WFC3IR camera. Assuming a similar performance as the one derived in this paper for GeMS/GSAOI, they conclude that the main limitation of GeMS/GSAOI compared to these other facilities is the multi-epoch noise floor, estimated to be around 0.4 mas, and which is a factor of more than two higher than for HST, and Keck NIRC2, both estimated to be $\sim$ 0.15 mas. The authors however emphasize that GeMS/GSAOI being the newer instrument, improvement of its astrometric properties, and in particular a better characterization of the potential time-variable distortions, is still under development and performance may improve as the system is being used. In addition, hardware solutions like the diffraction grid proposed by \cite{guyon2012astrometry, bendek2012astrometry, ammons2013astrometry} may improve consequently the potential astrometric performance. 

\section{Conclusion}
\label{sec:conclusion2}

We have presented a detailed analysis of the GeMS/GSAOI astrometric performance on crowded stellar fields. We show that the GeMS/GSAOI system has a large amount of high-order and time-variable distortion. The large distortions are mainly due to the AO bench, which uses an optical design with two off-axis parabolas. Moreover, with the AO-systems mounted on the Cassegrain focus of the Gemini telescope, changes in the gravity vector likely result in beam wander, which introduces time-variations in the distortion pattern seen on the science camera. As a result, the astrometric performance in crowded stellar fields greatly improves when high-order transformations are fitted and removed from the images, both for single and for multi-epoch data. Of course, every degree of freedom used in the transformation is equivalent to information lost from the proper motion system. For single epoch data sets, an astrometric error of $\sim$150 $\mu$as can be reached by allowing 60 degrees of freedom in the transformation between images with exposure times exceeding one minute. For bright stars, the remaining error approximately matches that predicted by simulation of MCAO's spatially- and time-variable PSFs. A careful modeling and estimation of the PSFs should allow further improvements.

For multi-epoch data sets, a systematic noise floor of $\sim$0.4 mas appears to be the limiting factor for GeMS/GSAOI astrometric performance. This noise floor could be reduced to 0.3 mas if one restricts the analysis to the brightest 50\%  of stars, and a FoV of 30\arcsec x 30\arcsec. But this term remains a factor of two larger than the single-epoch precision. Further characterizations, calibrations, and methods to reduce this noise floor are under development, and will be presented in a companion paper (Ammons et al. in prep.). 

In terms of transfer of experience, the impact of large distortions in the science focal plane was recognized early by the NFIRAOS (the TMT MCAO system) design team and the MAORY (E-ELT MCAO system) team. The NFIRAOS team has opted for a four parabola optical relay system that has nearly zero distortions in the science path \citep{herriot2012, schoeck2013tmt}. Both instruments will also be mounted on a gravity invariant Nasmyth platform, hence astrometric performance should be more stable. 

Nevertheless, and even at the current performance level, astrometric precisions of $<$ 0.5 mas over the full GSAOI 85" field of view can enable many new experiments in astrometry studies of crowded stellar fields that have not been efficient or even possible with existing ground-based AO systems due to their limited fields of view.

\section*{Acknowledgments}

This work is based on observations obtained at the Gemini Observatory, which is operated by the 
Association of Universities for Research in Astronomy, Inc., under a cooperative agreement with the NSF on behalf of the Gemini partnership: the National Science Foundation (United States), the National Research Council (Canada), CONICYT (Chile), the Australian Research Council (Australia), Minist\'{e}rio da Ci\^{e}ncia, Tecnologia e Inova\c{c}\~{a}o (Brazil) and Ministerio de Ciencia, Tecnolog\'{i}a e Innovaci\'{o}n Productiva (Argentina).
Part of this work has been funded by the French ANR program WASABI - ANR-13-PDOC-0006-01.
J.R.~acknowledges support from the National Science Foundation (AST- 1102791).
S.M.A.~acknowledges the U.S. Department of Energy by Lawrence Livermore National Laboratory support under Contract DE-AC52-07NA27344.

\bibliographystyle{mn2e}

\bibliography{ao,gems}

\begin{thebibliography}{}

\bibitem[\protect\citeauthoryear{Ammons, Bendek, Guyon, Marois, Neichel,
  Galicher \& Macintosh}{Ammons et~al.}{2013}]{ammons2013astrometry}
Ammons M.,  Bendek E.,  Guyon O.,  Marois O.,  Neichel B.,  Galicher R.,
  Macintosh B.,  2013, in AO4ELT3 On-sky pathfinder tests of calibrated mcao
  astrometry and implications for mcao on elts

\bibitem[\protect\citeauthoryear{Beckers}{Beckers}{1988}]{beckers1988increasing}
Beckers J.~M.,  1988, in Hulrich M.-H.,  ed., Very large telecopes and their
  instrumentation Increasing the size of the isoplanatic patch size with
  multiconjugate adaptive optics.
p.~693

\bibitem[\protect\citeauthoryear{Bendek, Ammons, Belikov, Pluzhnik \&
  Guyon}{Bendek et~al.}{2012}]{bendek2012astrometry}
Bendek E.,  Ammons S.,  Belikov R.,  Pluzhnik E.,    Guyon O.,  2012, in
  Proceedings of SPIE Vol.~8442, High precision astrometry laboratory
  demonstration for exoplanet detection using a diffractive pupil telescope.
pp 844243--844243

\bibitem[\protect\citeauthoryear{Bertin \& Arnouts}{Bertin \&
  Arnouts}{1996}]{bertin1996}
Bertin E.,  Arnouts S.,  1996, A\&AS, 117, 394

\bibitem[\protect\citeauthoryear{Cameron, Britton \& Kulkarni}{Cameron
  et~al.}{2009}]{cameron2009}
Cameron P.~B.,  Britton M.~C.,    Kulkarni S.~R.,  2009, The Astronomical
  Journal, 137, 83

\bibitem[\protect\citeauthoryear{Carrasco, Edwards, McGregor, Winge, Young,
  Doolan, van Harmelen, Rigaut et~al.,}{Carrasco
  et~al.}{2012}]{carrasco2012results}
Carrasco E.~R.,  Edwards M.~L.,  McGregor P.~J.,  Winge C.,  Young P.~J.,
  Doolan M.~C.,  van Harmelen J.,  Rigaut F.~J.,    et~al., 2012, in
  Proceedings of SPIE Vol.~8447, Results from the commissioning of the {G}emini
  {S}outh adaptive optics imager ({GSAOI}) at {G}emini {S}outh {O}bservatory

\bibitem[\protect\citeauthoryear{Clarkson, Ghez, Morris, Lu, Stolte, McCrady,
  Do \& Yelda}{Clarkson et~al.}{2012}]{clarkson2012}
Clarkson W.~I.,  Ghez A.~M.,  Morris M.~R.,  Lu J.~R.,  Stolte A.,  McCrady N.,
   Do T.,    Yelda S.,  2012, The Astrophysical Journal, 751, 132

\bibitem[\protect\citeauthoryear{Diolaiti, Bendinelli, Bonaccini, Close \& et
  al.}{Diolaiti et~al.}{2000}]{diolaiti2000}
Diolaiti E.,  Bendinelli O.,  Bonaccini D.,  Close L.,    et al. 2000, A\&AS,
  147, 335

\bibitem[\protect\citeauthoryear{Ellerbroek}{Ellerbroek}{1994}]{ellerbroek1994first}
Ellerbroek B.~L.,  1994, JOSA A, 11, 783

\bibitem[\protect\citeauthoryear{Fried \& Belsher}{Fried \&
  Belsher}{1994}]{fried1994analysis}
Fried D.~L.,  Belsher J.~F.,  1994, JOSA A, 11, 277

\bibitem[\protect\citeauthoryear{Fritz, Gillessen, Trippe, Ott \& et al.}{Fritz
  et~al.}{2010}]{fritz2010}
Fritz T.,  Gillessen S.,  Trippe S.,  Ott T.,    et al. 2010, MNRAS, 401, 1177

\bibitem[\protect\citeauthoryear{Genzel, Sch\"odel, Ott, Eisenhauer \& et
  al.}{Genzel et~al.}{2003}]{genzel2003}
Genzel R.,  Sch\"odel R.,  Ott T.,  Eisenhauer F.,    et al. 2003, The
  Astrophysical Journal, 594, 812

\bibitem[\protect\citeauthoryear{Ghez, Salim, Weinberg, Lu, Do, Dunn, Matthews,
  Morris et~al.,}{Ghez et~al.}{2008}]{ghez2008gc}
Ghez A.~M.,  Salim S.,  Weinberg N.~N.,  Lu J.,  Do T.,  Dunn J.~K.,  Matthews
  K.,  Morris M.~R.,    et~al., 2008, The Astrophysical Journal, 689, 1044

\bibitem[\protect\citeauthoryear{Gilles, Correia, V\'eran, Wang \&
  Ellerbroek}{Gilles et~al.}{2012}]{gilles2012psf}
Gilles L.,  Correia C.,  V\'eran J.-P.,  Wang L.,    Ellerbroek B.,  2012,
  Applied Optics, 51, 7443

\bibitem[\protect\citeauthoryear{Gillessen, Eisenhauer, Trippe, Alexander,
  Genzel, Martins \& Ott}{Gillessen et~al.}{2009}]{Gillessen2009}
Gillessen S.,  Eisenhauer F.,  Trippe S.,  Alexander T.,  Genzel R.,  Martins
  F.,    Ott T.,  2009, The Astrophysical Journal, 692, 1075

\bibitem[\protect\citeauthoryear{Guyon, Bendek, Eisner, Angel, Woolf, Milster,
  Ammons, Shao et~al.,}{Guyon et~al.}{2012}]{guyon2012astrometry}
Guyon O.,  Bendek E.,  Eisner J.,  Angel R.,  Woolf N.~J.,  Milster T.~D.,
  Ammons S.~M.,  Shao M.,    et~al., 2012, The Astrophysical Journal
  Supplement, 200, 11

\bibitem[\protect\citeauthoryear{Herriot, Andersen, Atwood \& Byrnes}{Herriot
  et~al.}{2012}]{herriot2012}
Herriot G.,  Andersen D.,  Atwood J.,    Byrnes P.,  2012, in Proceeding of
  SPIE Vol.~8447, Tmt nfiraos: adaptive optics system for the thirty meter
  telescope.
p. 84471M

\bibitem[\protect\citeauthoryear{Johnston \& Welsh}{Johnston \&
  Welsh}{1994}]{johnston1994analysis}
Johnston D.~C.,  Welsh B.~M.,  1994, JOSA A, 11, 394

\bibitem[\protect\citeauthoryear{Jolissaint, Neyman, Christou \&
  Wizinowich}{Jolissaint et~al.}{2012}]{jolissaint2012psf}
Jolissaint L.,  Neyman C.,  Christou J.,    Wizinowich P.,  2012, in
  Proceedings of SPIE Vol.~8447, Adaptive optics point spread function
  reconstruction project at w. m. keck observatory: first results with faint
  natural guide stars.
pp 844728--844728

\bibitem[\protect\citeauthoryear{Kudryavtseva, Brandner, Gennaro, Rochau,
  Stolte, Andersen, Da~Rio, Henning, Tognelli, Hogg, Clark \&
  Waters}{Kudryavtseva et~al.}{2012}]{kud2012}
Kudryavtseva N.,  Brandner W.,  Gennaro M.,  Rochau B.,  Stolte A.,  Andersen
  M.,  Da~Rio N.,  Henning T.,  Tognelli E.,  Hogg D.,  Clark S.,    Waters R.,
   2012, ApJL, 750, L44

\bibitem[\protect\citeauthoryear{LeLouarn \& Tallon}{LeLouarn \&
  Tallon}{2002}]{miska2001tomo}
LeLouarn M.,  Tallon M.,  2002, JOSA-A, 19, 912

\bibitem[\protect\citeauthoryear{Lindegren}{Lindegren}{1978}]{lindegren1978}
Lindegren L.,  1978, in Modern astrometry; Proceedings of the Colloquium
  Photoelectric astrometry - a comparison of methods for precise image
  location.
p.~197

\bibitem[\protect\citeauthoryear{Lu, Neichel, Anderson, Sinukoff, Hosek, Ghez,
  Morris \& Rigaut}{Lu et~al.}{2014}]{Jessica2014spie}
Lu J.,  Neichel B.,  Anderson J.,  Sinukoff E.,  Hosek M.,  Ghez A.,  Morris
  M.,    Rigaut F.,  2014, in SPIE Vol. 9148-191, Near-infrared astrometry of
  star clusters with different flavors of adaptive optics and hst

\bibitem[\protect\citeauthoryear{Lu, Ghez, Hornstein, Morris, Becklin \&
  Matthews}{Lu et~al.}{2009}]{lu2009}
Lu J.~R.,  Ghez A.~M.,  Hornstein S.~D.,  Morris M.~R.,  Becklin E.~E.,
  Matthews K.,  2009, The Astrophysical Journal, 690, 1463

\bibitem[\protect\citeauthoryear{McGregor, Hart, Stevanovic, Bloxham, Jones,
  Van~Harmelen, Griesbach, Dawson et~al.,}{McGregor
  et~al.}{2004}]{mcgregor2004gemini}
McGregor P.,  Hart J.,  Stevanovic D.,  Bloxham G.,  Jones D.,  Van~Harmelen
  J.,  Griesbach J.,  Dawson M.,    et~al., 2004, in Proceeding of SPIE
  Vol.~5492, Gemini south adaptive optics imager (gsaoi).
pp 1033--1044

\bibitem[\protect\citeauthoryear{Marchetti, Brast, Delabre, Donaldson, Fedrigo,
  Frank, Hubin, Kolb et~al.,}{Marchetti et~al.}{2007}]{marchetti2007mad}
Marchetti E.,  Brast R.,  Delabre B.,  Donaldson R.,  Fedrigo E.,  Frank C.,
  Hubin N.,  Kolb J.,    et~al., 2007, in Adaptive Optics: Methods, Analysis
  and Applications Mad on-sky results in star oriented mode

\bibitem[\protect\citeauthoryear{Meyer, K\"urster, Arcidiacono, Ragazzoni \&
  Rix}{Meyer et~al.}{2011}]{Meyer2011}
Meyer E.,  K\"urster M.,  Arcidiacono C.,  Ragazzoni R.,    Rix H.-W.,  2011,
  Astronomy and Astrophysics, 532, 10

\bibitem[\protect\citeauthoryear{Munro \& Dubois}{Munro \&
  Dubois}{1995}]{munro1995yorick}
Munro D.~H.,  Dubois P.~F.,  1995, Computers in Physics, 9, 609

\bibitem[\protect\citeauthoryear{Neichel, Rigaut, Vidal \& et al.}{Neichel
  et~al.}{2014}]{neichel2014gems}
Neichel B.,  Rigaut F.,  Vidal F.,    et al. 2014, MNRAS, 440, 1002

\bibitem[\protect\citeauthoryear{Rigaut}{Rigaut}{2014}]{Rigaut2014ngs2}
Rigaut F.,  2014, in SPIE Vol. 9148-191, Ngs2: the natural guide star next
  generation sensor for gems

\bibitem[\protect\citeauthoryear{Rigaut, Neichel, Bec \&
  Garcia-Rissmann}{Rigaut et~al.}{2010}]{rigaut2010myst}
Rigaut F.,  Neichel B.,  Bec M.,    Garcia-Rissmann A.,  2010, in Proceedings
  of SPIE Vol.~7736, Myst: a comprehensive high-level {AO} control tool for
  {GeMS}

\bibitem[\protect\citeauthoryear{Rigaut, Neichel, Boccas, d'Orgeville,
  Arriagada, Fesquet, Diggs, Marchant et~al.,}{Rigaut
  et~al.}{2012}]{rigaut2012gems}
Rigaut F.,  Neichel B.,  Boccas M.,  d'Orgeville C.,  Arriagada G.,  Fesquet
  V.,  Diggs S.~J.,  Marchant C.,    et~al., 2012, in Proceedings of SPIE
  Vol.~8447, Gems: First on-sky results

\bibitem[\protect\citeauthoryear{Rigaut, Neichel, Boccas \& et al.}{Rigaut
  et~al.}{2014}]{rigaut2014gems}
Rigaut F.,  Neichel B.,  Boccas M.,    et al. 2014, MNRAS, 437, 2361

\bibitem[\protect\citeauthoryear{Rochau, Brandner, Stolte, Gennaro, Gouliermis,
  Da~Rio, Dzyurkevich \& Henning}{Rochau et~al.}{2010}]{Rochau2010}
Rochau B.,  Brandner W.,  Stolte A.,  Gennaro M.,  Gouliermis D.,  Da~Rio N.,
  Dzyurkevich N.,    Henning T.,  2010, ApJL, 716, L90

\bibitem[\protect\citeauthoryear{Sch\"odel}{Sch\"odel}{2010}]{schodel2010}
Sch\"odel R.,  2010, A\&A, 509, 16

\bibitem[\protect\citeauthoryear{Schoeck, Do, Ellerbroek, Herriot, Meyer,
  Suzuki, Wang \& Yelda}{Schoeck et~al.}{2013}]{schoeck2013tmt}
Schoeck M.,  Do T.,  Ellerbroek B.,  Herriot G.,  Meyer L.,  Suzuki R.,  Wang
  L.,    Yelda S.,  2013, in AO4ELT3 Developing performance estimates for high
  precision astrometry with tmt

\bibitem[\protect\citeauthoryear{Stetson}{Stetson}{1987}]{stetson1987}
Stetson P.~B.,  1987, The Publications of the Astronomical Society of the
  Pacific, 99, 191

\bibitem[\protect\citeauthoryear{Stolte, Ghez, Morris, Lu, Brandner \&
  Matthews}{Stolte et~al.}{2008}]{stolte2008}
Stolte A.,  Ghez A.~M.,  Morris M.,  Lu J.~R.,  Brandner W.,    Matthews K.,
  2008, The Astrophysical Journal, 675, 1278

\bibitem[\protect\citeauthoryear{Tallon \& Foy}{Tallon \&
  Foy}{1990}]{tallon1990adaptive}
Tallon M.,  Foy R.,  1990, Astronomy and Astrophysics, 235, 549

\bibitem[\protect\citeauthoryear{Trippe, Davies, Eisenhauer, Forster-Schreiber
  \& et al.}{Trippe et~al.}{2010}]{trippe2010}
Trippe S.,  Davies R.,  Eisenhauer F.,  Forster-Schreiber N.~M.,    et al.
  2010, MNRAS, 402, 1126

\bibitem[\protect\citeauthoryear{Yelda, Ghez, Lu, Do, Meyer, Morris \&
  Matthews}{Yelda et~al.}{2014}]{yelda2014astrometry}
Yelda S.,  Ghez A.~M.,  Lu J.~R.,  Do T.,  Meyer L.,  Morris M.~R.,    Matthews
  K.,  2014, The Astrophysical Journal, 783, 131

\bibitem[\protect\citeauthoryear{Yelda, Lu, Ghez, Clarkson \& et al.}{Yelda
  et~al.}{2010}]{yelda2010astrometry}
Yelda S.,  Lu J.~R.,  Ghez A.~M.,  Clarkson W.,    et al. 2010, The
  Astrophysical Journal, 725, 331

\end{thebibliography}

\label{lastpage2}

\end{document}